\documentclass{appolb}
\usepackage{epsfig}

\usepackage{graphicx}
\usepackage{amsmath}
\usepackage{amssymb}
\newcommand{\eq}{\begin{equation}}
\newcommand{\qe}{\end{equation}}
\newcommand{\eqa}{\begin{eqnarray*}}
\newcommand{\qa}{\end{eqnarray*}}

\newcommand{\ho}{\hspace{.1cm}}

\begin{document}
\title{\mbox{\boldmath $S$-, $P$- and $D$-wave  \mbox{\boldmath $\pi \pi$} final state interactions
 and $CP$} violation in \mbox{\boldmath $B^\pm \to \pi^\pm \pi^\mp \pi^\pm$} decays}

\author{J.-P.~Dedonder and B.~Loiseau
\address{Laboratoire de Physique Nucl\'eaire et de Hautes \'Energies, Groupe Th\'eorie, \\
Universit\'e Pierre et Marie Curie et Universit\'e Paris-Diderot, IN2P3 \& CNRS, 4~place Jussieu, 75252 Paris, France}
\and
A.~Furman
\address{ul. Bronowicka 85/26, 30-091 Krak\'ow, Poland}
\and
R.~Kami\'nski and L.~Le\'sniak
\address{Division of Theoretical Physics, The Henryk Niewodnicza\'nski Institute of Nuclear Physics, Polish Academy of Sciences, 31-342 Krak\'ow, Poland}
}
\maketitle
\begin{abstract}

We study $CP$ violation and the contribution of the strong pion-pion interactions in the three-body $B^\pm\to \pi^\pm \pi^\mp \pi^\pm$ decays within a quasi two-body QCD factorization approach.
The short distance interaction amplitude is calculated in the next-to-leading order in the strong coupling constant with vertex and penguin corrections.
The meson-meson final state interactions are described by pion non-strange  scalar and vector form factors for the $S$ and $P$ waves and by a relativistic Breit-Wigner formula for the $D$ wave.
The pion scalar form factor is calculated from a unitary relativistic  coupled-channel model including  $\pi \pi$, $K \bar K$ and effective $(2\pi)(2\pi)$ interactions.
The pion vector form factor results from a Belle Collaboration analysis of  $\tau^- \to \pi^- \pi^0 \nu_\tau$ data.
The recent $B^\pm\to \pi^\pm \pi^\mp \pi^\pm$  BABAR Collaboration data are fitted with our model using only three parameters for the $S$ wave, one for the $P$ wave and one for the $D$ wave.
We find not only a sizable contribution of the $S$ wave just above the $\pi \pi$ threshold but also under the $\rho(770)$ peak a significant interference, mainly between the $S$ and $P$ waves. 
For the $B$ to $f_2(1270)$ transition form factor, we predict $F�^{Bf_2}(m_\pi^2)=0.098\pm0.007$.
Our model yields a unified unitary description of the contribution of the three scalar 
 resonances $f_0(600)$, $f_0(980)$ and $f_0(1400)$ in terms of the pion non-strange scalar form factor.

\end{abstract}
  
\section{Introduction}
\label{introduction}

Three-body charmless hadronic $B$ meson decays offer one of the best tools for studies of direct $CP$ violation and provide an interesting testing ground  for  strong interaction dynamical models.
The present work, part of a program devoted to the understanding of rare three-body $B$ decays~\cite{fkll,El-Bennich2006,Bppk, Leitner_PRD81},  
 is motivated by the recent BABAR Dalitz-plot analysis of the $B^\pm \to \pi^\pm \pi^\mp \pi^\pm$ decays~\cite{Aubert:2009}.
In an isobar model description, the authors of Ref.~\cite{Aubert:2009} find evidence for the $f_0(1370)$ but, within the current experimental accuracy,  no significant signal for the $f_0(980)$. The $f_0(600)$, not explicitly included in that analysis, could be part of the non-resonant background.
Furthermore, there is a small but visible contribution of the $f_2(1270)$ resonance~\cite{Aubert:2009}.
 
Here, the aim is to provide a phenomenological analysis of the $B^\pm \to \pi^\pm \pi^\mp \pi^\pm$ decay channels relying on the QCD factorization scheme (QCDF) in the  $\pi \pi$ effective mass range from threshold to 1.64~GeV.
The focus will be set on the final state $\pi \pi$ interactions involved since a partial wave analysis of the Dalitz plot should use theoretically and phenomenologically well constrained  $\pi \pi$ amplitudes. 

Studies of $B$ decays into two-body and quasi-two-body final states have been performed in the QCDF framework~\cite{Ali1998,Beneke:2001ev,Gardner_PRD65,bene03,LeitnerQCDF,Cheng:2005nb,Cheng0704.1049}. 
The naive factorization approach is a useful first order approximation which receives corrections proportional to the strong coupling constant $\alpha_s$ at scales $m_b$ and  $\sqrt{\Lambda_{QCD} m_b}$ and in inverse powers of the $b$ quark mass $m_b$~\cite{Beneke2006}. 
In the present study, we propose an extension of these results to the three-body decays $B^\pm\to \pi^\pm \  \pi^+ \pi^-$.

The role of the $f_0(600)$ (or $\sigma$) in  charmless three-body decays of $B$ mesons has been examined by Gardner and Mei\ss ner~\cite{Gardner_PRD65} in  $B^0 \to \pi^+ \pi^- \pi^0$ decays.
Within  QCD quasi two-body factorization approach their $f_0(600)  \pi$ amplitude is described by a unitary pion scalar form factor constrained by $\pi \pi $ scattering and chiral dynamics.
This is different from the relativistic Breit-Wigner parametrization  used in most experimental analyses and in some theoretical studies, for example in~\cite{Deandrea_PRL86}. 
This has led to improved theoretical predictions; the contribution of the  $f_0(600) \pi$  channel has been found to be important in the range of the dominant $\rho^0 \pi^0$ intermediate state. 
However, in recent $B^0 \to \pi^+ \pi^- \pi^0$  Dalitz plot analyses~\cite{BABAR_PRD76, Belle_PRD77} no contribution from $B^0 \to f_0(600) \pi^0$ channel has been found. 
This could be linked to the present limited statistics in the low effective $\pi \pi$ mass region.
Furthermore, such a contribution could be hidden in the nonresonant amplitude introduced in the experimental analysis.
Nevertheless we will show that the contribution of the $S$ wave is important in the  $B^{\pm} \to \pi^{\pm} \ \pi^+ \pi^-$ decays.

Charmless three-body decays of $B$ mesons have also been investigated by Cheng, Chua and Soni~\cite{Cheng0704.1049}  in the framework of quasi two-body factorization approach using resonant and non-resonant contributions. 
In particular they have calculated the $B^- \to \pi^+ \pi^- \pi^-$ branching fractions and $CP$ asymmetries and found a small rate for $B^- \to f_0(980) \pi^-$decay.

An achievement in the theory of $B$ decays into two mesons is the confirmation of the validity of factorization as a leading order  approximation.
No proof of factorization has yet been given for the $B$ decays into three mesons. However, three-body interactions are suppressed when specific kinematical configurations with the three mesons  quasi aligned in the rest frame of the $B$ meson are considered. 
This is the case in the effective $\pi^+ \pi^-$ mass region smaller than 1.64~GeV in the Dalitz plot where most of the $\pi^+ \pi^-$ resonant states are visible. 
Such processes will be denoted as $B^\pm \to \pi^\pm [\pi^+\pi^-]$, the mesons of the  [$\pi^+\pi^-$] pair moving more or less, in the same direction in the $B$ rest frame.
Then, it seems reasonable to postulate the validity of factorization for this quasi two-body $B$ decay~\cite{Beneke3body} assuming that the [$\pi^+ \pi^-$] pair originates from a quark-antiquark state.

In the factorization approach the $B^{\pm} \to \pi^{\pm}_1 \ [\pi^+_2 \pi^-_3]$ decay amplitudes are expressed as a superposition of appropriate  effective QCD coefficients  and two products of two transition matrix elements.
The transition matrix elements  between the $B^{\pm}$ meson and the $\pi^{\pm}_1$   pion  multiplied by the transition matrix elements between the vacuum and the $\left[\pi^+_2 \pi^-_3\right]$ pion pair  correspond to the first of these products.
Here, in the $\pi^+_2 \pi^-_3$ center of mass frame, the bilinear quark currents involved force the $[\pi^+_2 \pi^-_3]$
 pair to be in $S$ or in $P$ state. 
The second term is associated to the product of the transition matrix elements between the $B^{\pm}$ meson 
and the $[\pi^+_2 \pi^-_3]$ pion pair in $S$, $P$ or $D$ state by the transition matrix elements between the vacuum and the $\pi^{\pm}_1$ pion.
The $[\pi^+_2 \pi^-_3]_{S,P}$ transition matrix elements to the vacuum are proportional to the pion scalar and vector form factors. 
We assume that the $B^{\pm} \to \pi^+_2 \pi^-_3$ matrix elements are expressed as products of the $B^{\pm} \to [\pi^+_2 \pi^-_3]_{S,P,D}$ transition form factors by the relevant vertex function describing the decay of the $ [\pi^+_2 \pi^-_3]_{S,P,D}$ state into the final pion pair. 
The vertex functions are in turn assumed to be proportional to the pion scalar form factor  for the $S$ wave, to the vector form factor for the $P$ wave and to a relativistic Breit-Wigner formula for the $D$ wave.
Here, a single unitary function, namely the pion non-strange scalar form factor,  describes then the three scalar resonances, $f_0(600)$, $f_0(980)$ and $f_0(1400)$ present in the  $\pi^+ \pi^-$ interaction.

In Sec.~\ref{Model} we present the model used in the analysis. 
Sec.~\ref{SVff} is devoted to the construction of the pion scalar and vector form factors. 
The pertinent observables and the fitting procedure are described in Sec.~\ref{fit} while the results are
 discussed in Sec.~\ref{results}. 
A summary and some perspectives are outlined in the final Sec.~\ref{Outlook}. 
The detailed derivation of the decay amplitudes is presented in the Appendix~\ref{appendix} while Appendix~\ref{alpha_i_equation} gives the system of equations to be solved to obtain the parameters fixing the low-energy behavior of the pion scalar form factor to be that of one loop calculation in chiral perturbation theory.

\section{Decay amplitudes}
\label{Model}
The amplitudes for the non-leptonic decays of the $B$ meson 
are given as matrix elements of the effective weak Hamiltonian~\cite{Ali1998,Beneke:2001ev}

\begin{equation}
\label{Heff}
H_{eff}=\frac{G_F}{\sqrt{2}} \sum_{p=u,c}\lambda_p\, \Big[ C_1 O_1^p + C_2 O_2^p
+\sum_{i=3}^{10} C_i O_i + C_{7\gamma} O_{7\gamma} + C_{8g} O_{8g} \Big] + h.c., 
\end{equation}
where  

\begin{equation}
\label{lambda}
\lambda_u= V_{ub} V^*_{ud},\ 
\lambda_c= V_{cb} V^*_{cd},
\end{equation}
the $V_{pp'}\ (p'=b,d)$ being Cabibbo-Kobayashi-Maskawa quark-mixing matrix elements.
For the Fermi coupling constant $G_F$ we take the value $1.16637 \times 10^{-5}$ GeV$^{-2}$. 
The $C_i(\mu)$ are the Wilson coefficients of the  four-quark operators  $O_i(\mu)$ at a renormalization scale $\mu$.
The $O_{1,2}^p$ are left-handed current-current operators arising from $W$-boson exchange, $O_{i=3-10}$ are QCD and electroweak penguin operators involving a loop with a $u$ or $c$ quark and a $W$ boson, $O_{7\gamma}$ and $O_{8g}$ are
the electromagnetic and chromomagnetic dipole operators~\cite{Beneke:2001ev}.

Let $p_B$ be the four-momentum of the $B^{\pm}$ meson and $p_1$ that of the isolated $\pi^{\pm}$. 
Let  then  $p_2$  denote the four-momentum of the $\pi^+$ and  $p_3$  that of the $\pi^-$ of the interacting [$\pi^+\pi^-$] pair in the $B$ rest frame.
One has $p_B=p_1+p_2+p_3$ and we introduce the invariants $s_{ij}= (p_i + p_j)^2$ for $i,j = 1,2,3$ with $i < j$. 
For  the $B^-\to \pi^- \ [\pi^+ \pi^-]_{{S},{P},D}$ amplitude, we work in the  center of mass frame of the $\pi^+ \pi^-$ pair of pions with respective four-momenta $p_2$  and $p_3$ (or $p_1$  and $p_2$ for the symmetrized amplitudes). 
These two pions will be either in a relative $S$, $P$ or $D$ state.
In the following we derive the amplitudes for the $B^-\to \pi^- \ho [\pi^+ \pi^-]_{{S},{P},D}$ processes.  
The transcription to the $B^+\to \pi^+ \ho [\pi^+ \pi^-]_{{S},{P},D}$ processes  is straightforward. 
Applying the QCD factorization formula for the $B^-\to \pi^- \ [\pi^+ \pi^-]_{{S},{P},D}$ process, the matrix elements of the effective weak Hamiltonian~(\ref{Heff}) can be written as~\cite{Beneke:2001ev}
\begin{eqnarray}
\label{3piTpB}
\left \langle \pi^-(p_1)\ [\pi^+(p_2){\pi^-}(p_3) ]_{{S},{P},D}\vert H_{eff}\vert B^-(p_B) \right \rangle = \nonumber\\
\frac{G_F}{\sqrt{2}}\ho  \sum_{p=u,c} \lambda_p \ho \left \langle  \pi^- \ [\pi^+{\pi^-}]_{{S},{P},D}\vert T_p \vert B^-\right \rangle,
\end{eqnarray}
to which must be added the symmetrized term \\ $\langle \pi^-(p_3) [\pi^+(p_2){\pi^-}(p_1)]_{{S},{P},D} \vert H_{eff}\vert B^-(p_B) \rangle $. 
With $M_1\equiv \pi^-$ and $M_2\equiv [\pi^+\pi^-]_{{S},{P}}$ or $M_1\equiv [\pi^+\pi^-]_{{S},{P},D} $ while $M_2\equiv \pi^-$,  one has
  
\begin{eqnarray}
\label{Tp}
\left \langle  \pi^- \ [\pi^+{\pi^-}]_{{S},{P},D}\vert T_p\vert B^-\right \rangle&=&  \langle  \pi^- \ [\pi^+{\pi^-}]_{{S},{P},D}\vert \nonumber\\
        &\ho\Big\{ &a_1(M_1M_2)\delta_{pu} (\bar u b)_{V-A} \otimes (\bar d u)_{V-A}\nonumber\\
	&+&a_2(M_1M_2)\delta_{pu} (\bar d b)_{V-A} \otimes (\bar u u)_{V-A} \nonumber\\
	&+&a_3(M_1M_2)\sum_q(\bar d b)_{V-A} \otimes (\bar q q)_{V-A} \nonumber\\
	&+&a_4^p(M_1M_2)\sum_q(\bar q b)_{V-A} \otimes (\bar d q)_{V-A} \nonumber\\
	&+&a_5(M_1M_2)\sum_q(\bar d b)_{V-A} \otimes (\bar d q)_{V+A} \nonumber\\
	&+&a_6^p(M_1M_2)\sum_q(-2)(\bar q b)_{sc-ps} \otimes (\bar d q)_{sc+ps} \nonumber\\ 
	&+&a_7(M_1M_2)\sum_q(\bar d b)_{V-A} \otimes \frac{3}{2}e_q (\bar q q)_{V+A} \nonumber\\
	&+&a_8^p(M_1M_2)\sum_q(-2)(\bar q b)_{sc-ps} \otimes \frac{3}{2}e_q (\bar d q)_{sc+ps}
	 \nonumber\\
	&+&a_{9}(M_1M_2)\sum_q(\bar d b)_{V-A} \otimes \frac{3}{2}e_q (\bar q q)_{V-A}\nonumber\\
	&+&a_{10}^p(M_1M_2)\sum_q(\bar q b)_{V-A} \otimes \frac{3}{2}e_q (\bar d q)_{V-A}\Big\}\ \nonumber\\
        &\vert&B^- \rangle,
\end{eqnarray}
 where $a_j^{p}$ are effective QCDF coefficients.\\ 
 In Eq.(\ref{Tp}),  $(\bar q_{1}q_{2})_{V \mp A} = \bar q_{1} \gamma_\mu(1\mp \gamma_5)q_{2}$, $(\bar q_1 q_2)_{sc\pm ps}=\bar q_1(1\pm \gamma_5)q_2$ and \\ $e_q$ denotes the electric charge of the quark $q$ in  units of the elementary charge $e$. 
The sum on the index $q$ runs over $u$ and $d$
and the summation over the color degree of freedom has been performed. 
The notations $sc$ and $ps$ stand for scalar and pseudoscalar, respectively.
 
 At next-to-leading order (NLO) in the strong coupling constant $\alpha_s$,  the general expression 
of the $a_i^p$ quantities  in terms of effective Wilson coeffficients is~\cite{bene03}
 
\begin{eqnarray}
 \label{aip}
 a_i^p(M_1M_2)=\left(C_i+\frac{C_{i\pm 1}}{N_C}\right) N_i(M_2)+ 
\frac{C_{i\pm 1}}{N_C}\frac{C_F \alpha_s}{4\pi}
 \left [V_i(M_2)
\right.
\nonumber\\
\left.
+\frac{4\pi^2}{N_C}H_i(M_1M_2)\right]+P_i^p(M_2),
\end{eqnarray}
 where the upper (lower) signs apply when the index $i$ is odd (even), $N_C$ is the number of
colors,  $N_C=3$  and $C_F=(N_C^2-1)/2N_C$. The sums over the color degree of freedom have been
 performed in Eq.~(\ref{Tp}). 
Note that in the leading-order (LO) contribution $N_i(M_2)=0$ for $M_2= [\pi^+\pi^-]_{P}$ and $i=6, 8$, otherwise  $N_i(M_2)=1$.
The NLO quantities $V_i(M_2)$ arise from one loop vertex corrections, $H_i(M_1M_2)$ from hard spectator
 scattering interactions and $P_i^p(M_2)$ from penguin contractions.
Here the meson $M_2$ is the meson which does not include the spectator quark of the $B$ meson.
The superscript $p$ in $a_i^p(M_1M_2)$ is to be omitted for $ i=1$, $2$, $3$, $5$, $7$ and $9$ since the penguin corrections are equal to zero in these cases.
In our calculation we shall not include the NLO hard scattering corrections nor the annihilation
contributions which require the introduction of four phenomenological parameters to regularize end point
 divergences  related to asymptotic wave functions~\cite{bene03}. Although we are aware that such 
contributions might be important, this would bring, at this stage of analysis, too many free parameters.

In Eq.~(\ref{Tp}) the symbol $\otimes$ indicates that the different components of the matrix elements $\left \langle  \pi^- \ [\pi^+{\pi^-}]_{{S},{P},D}\vert T_p \vert B^-\right \rangle$ are to be calculated in the factorized form,

\begin{eqnarray}
\label{j1facj2}
\left \langle  \pi^-(p_1) [\pi^+(p_2){\pi^-}(p_3) ]_{{S},{P},D}\  \vert j_1\otimes j_2\vert B^-(p_B)
 \right \rangle  \nonumber\\
\equiv 
 \left\langle [\pi^+\pi^-]_{{S},{P},D}\vert j_1\vert B^- \right \rangle
 \left \langle \pi^-\ \vert j_2\vert 0 \right \rangle\  {\rm or}
 \left \langle \pi^-\vert j_1\vert B^-\right \rangle  
 \left\langle [\pi^+\pi^-]_{{S},{P}} \ \vert j_2\vert 0  \right \rangle,
 \end{eqnarray}
since we neglect $B^-$ annihilation contributions which are expected to be small~\cite{Ali1998}. 
Furthermore, as for the hard scattering corrections, their evaluation~\cite{bene03} introduces two phenomenological parameters.
In Eq.~(\ref{j1facj2}) $j_1$ and $j_2$ denote the appropriate quark currents entering in Eq.~(\ref{Tp}).
Note that, in our approach, in the evaluation of the long distance matrix element $\left\langle [\pi^+\pi^-]_{{S},{P},D}\vert j_1\vert B^- \right \rangle$,  we make the hypothesis that the transitions of $B^-$ to the  $[\pi^+\pi^-]_{{S},{P},D}$ states go first through intermediate meson resonances  $R_{{S},{P},D}$  which then decay into a $\pi^+\pi^-$ pair.
We describe these decays by a vertex function  modeled by assuming them to be proportional to the pion scalar or vector form factors or to a relativistic Breit-Wigner formula, respectively.
For the short distance part of the decay amplitudes proportional to a combination of the effective coefficients $a_i^p(M_1 M_2)$ it can be seen that for terms coming from the first line of the right hand side of Eq.~(\ref{j1facj2}) $M_1\equiv [\pi^+ \pi^-]_{S,P,D}$,  $M_2\equiv \pi^-$ and for those from the second line  $M_1\equiv \pi^-$ while  $M_2\equiv [\pi^+ \pi^-]_{S,P}$, the $ [\pi^+ \pi^-]_D$ transition to the vacuum being zero with the involved bilinear quark current $j_2$ in Eq.~(\ref{j1facj2}) .
In the following, when $M_2\equiv [\pi^+ \pi^-]_{S,P}$, we assume that the NLO corrections $V_i(M_2)$ and $P_i^p(M_2)$ are evaluated at the meson resonances $R_{{S},{P}}$ position. 
Here we take $R_P\equiv \rho(770)^0$ and $R_S\equiv f_0(980)$.
A similar approximation has been applied in Refs.~\cite{Bppk, Leitner_PRD81} for the $[K\pi]_{{S},{P}}$ states with $R_P\equiv K^*(892)$ and  $R_S\equiv K_0^*(1430)$.

Introducing the following short distance terms, with $L \equiv S, P, D$ and with $R_D\equiv f_2(1270)$,

\begin{eqnarray}
\label{RSpi}
u(R_ {L} \pi^-)=\lambda_u\left
\{
a_1(R_ {L}\pi^-) +  a_4^u(R_{L}\pi^-)  +a_{10}^u(R_{L}\pi^-)-
\left[
 a_6^u(R_{L}\pi^-)+
\right.
\right.
 \nonumber\\
\left.
\left.
a_8^u(R_{L}\pi^-) 
\right]
r_\chi^\pi
\right \} 
+ \lambda_c
\left\{
    a_4^c(R_{L}\pi^-)  +a_{10}^c(R_{L}\pi^-)-
\left[
 a_6^c(R_{L}\pi^-)+
\right.
\right.
 \nonumber\\
\left. 
\left.
 a_8^c(R_{L}\pi^-)
\right]
  r_\chi^\pi 
\right\}
 ,
\end{eqnarray}
\begin{eqnarray}
\label{piRS}
v(\pi^- R_S)=\lambda_u  
\left 
[
-2a_6^u(\pi^-R_S)+a_8^u(\pi^-R_S)
\right
]
+ \lambda_c 
\left
[
-2a_6^c(\pi^-R_S)
\right.
\nonumber\\
\left.
+a_8^c(\pi^-R_S)
\right
]
, 
\end{eqnarray}
and 

 \begin{eqnarray}
\label{wRP}
w(\pi^- R_P)=\lambda_u \left\{
a_2(\pi^-R_P)
 -a_4^u(\pi^-R_P)+\frac{3}{2}\left [
a_7(\pi^-R_P)+a_9(\pi^-R_P)\right]
\right.
 \nonumber\\
\left. 
 +\frac{1}{2}a_{10}^u(\pi^-R_P)\right\}
\nonumber\\
+ \lambda_c \left \{
 -a_4^c(\pi^-R_P)+\frac{3}{2}\left [
a_7(\pi^-R_P)+a_9(\pi^-R_P)\right] 
 +\frac{1}{2}a_{10}^c(\pi^-R_P)\right \},
\end{eqnarray}
one obtains, from Eqs.~(\ref{3piTpB}), (\ref{Tp}) and (\ref{j1facj2}), the following $S$-,  
$P$- and $D$-wave matrix elements
 
\begin{equation}
\label{uv}
\sum_{p=u,c} \lambda_p \ho \left \langle  \pi^-(p_1)  [\pi^+(p_2){\pi^-}(p_3) ]_{S}\vert  T_p  \vert B^- \right \rangle 
=
X_S \ho u(R_S \pi^-) + Y_S \ho v(\pi^- R_S),
\end{equation}

\begin{equation}
\label{uw}
\sum_{p=u,c} \lambda_p \ho  \left \langle \pi^-(p_1)  [\pi^+(p_2){\pi^-}(p_3) ]_{P}\vert  T_p  \vert
 B^- \right \rangle 
=
 X_P \ho u(R_P \pi^-) + Y_P \ho w(\pi^- R_P),
\end{equation}

\begin{equation}
\label{u}
\sum_{p=u,c} \lambda_p \ho  \left \langle \pi^-(p_1)  [\pi^+(p_2){\pi^-}(p_3) ]_{D}\vert  
T_p  \vert B^- \right \rangle 
=X_D \ho u(R_D \pi^-).
\end{equation}
In Eq. (\ref{RSpi})  the chiral factor $r_\chi^\pi$ is given by  $r_\chi^\pi= 2m_\pi^2/[(m_b+m_u)(m_u+m_d)]$, $m_u$ and $m_d$ being the $u$ and $d$ quark masses, respectively. 
The long distance functions $X_{{S},{P},D}$ and $Y_{S,{P}}$, evaluated in Appendix~\ref{appendix}, read 

 \begin{eqnarray}
 \label{XYZ}
 \label{XS}
X_{{S}} &\equiv& \left \langle  [\pi^+(p_2){\pi^-}(p_3) ]_{S} \vert (\bar u b)_{V-A}\vert B^-\right \rangle \ho \left \langle \pi^-(p_1)\vert  (\bar d u)_{V-A}\vert 0 \right \rangle \nonumber\\
&=&   -\sqrt{\frac{2}{3}} \ho \chi_S \ho f_\pi \ho (M_{B}^2-s_{23}) \ho F_0^{BR_S}(m_\pi^2) \ho\Gamma_1^{n*}(s_{23}),   \\
 \label{ZS}
Y_{{S}} &\equiv& \left \langle \pi^-(p_1)\vert (\bar d b)_{sc-ps}\vert B^- \right \rangle \ho \left \langle  [\pi^+(p_2){\pi^-}(p_3) ]_{S}\vert (\bar d d)_{sc+ps}\vert 0\right \rangle\nonumber \\
&=&  \ \sqrt{\frac{2}{3}} \ho B_0\ho \frac {M_B^2-m_\pi^2}{m_b-m_d}
\ho F_0^{B\pi}(s_{23}) \ho \Gamma_1^{n*}(s_{23}),  \\
 \label{XP}
X_{{P}} &\equiv & \left \langle  [\pi^+(p_2){\pi^-}(p_3) ]_{P}  \vert (\bar u b)_{V-A}\vert B^-\right \rangle \ho \left \langle \pi^-(p_1)\vert  (\bar d u)_{V-A}\vert 0 \right \rangle \nonumber \\
 &=&   N_P \frac{f_\pi}{f_{R_P}} \ho (s_{13} - s_{12}) \ho A_0^{B R_P}(m_\pi^2) \ho  F_1^{\pi\pi}(s_{23}), \\
 \label{YP}
Y_{{P}} &\equiv& \left \langle \pi^-(p_1)\vert  (\bar d b)_{V-A}\vert B^- \right \rangle \ho \left \langle  [\pi^+(p_2){\pi^-}(p_3) ]_{P}\vert (\bar u u)_{V-A} \vert 0 \right \rangle \nonumber \\
&=&     (s_{13} - s_{12}) F_1^{B\pi}(s_{23}) F_1^{\pi \pi}(s_{23}),\\
 \label{XD}
X_{{D}} &\equiv & \left \langle  [\pi^+(p_2){\pi^-}(p_3) ]_{D}  \vert (\bar u b)_{V-A}\vert B^-\right \rangle \ho \left \langle \pi^-(p_1)\vert  (\bar d u)_{V-A}\vert 0 \right \rangle \nonumber \\
 &=&  -\frac{f_\pi}{\sqrt2}   F^{B R_D}(m_\pi^2)  \ho \sqrt{\frac{2}{3}}\frac{G_{f_2} D(s_{12},s_{23})}{m^2_{R_D}-s_{23}-im_{R_D}\Gamma(s_{23})},
 \end{eqnarray}
The different quantities entering the above equations are discussed below.

The $S$-wave strength parameter $\chi_S$ [Eq. (\ref{XS})] will be fitted together with 
the correction $P$-wave parameter $N_P$ [Eq. (\ref{XP}]. 
The deviation of $N_P$ from 1 corresponds
to the possible variation of the strength of this $P$-wave amplitude proportional to $f_{\pi}/f_{R_P}$ 
[compare Eqs.~(\ref{RSvertex}) and (\ref{vertexRP})].

Three scalar-isoscalar $f_0$ resonances, viz. $f_0(600)$, $f_0(980)$ and $f_0(1400)$, are present 
in the $\pi\pi$ effective mass range, $m_{\pi\pi}$, considered here. 
Since some of them are wide, like $f_0(600)$, one could have a possible $R_S$
dependence in $\chi_S$.
The transition form factor from $B$ to $R_S$, $F_0^{BR_S}(m_\pi^2)$, could also depend on 
$m_{\pi\pi}$. 
However,  one expects these dependences to be weaker than the effective mass dependence of the pion
scalar form factor, $\Gamma_1^{n*}(s_{23})$, in which all these resonances are incorporated. 
Therefore we assume that $\chi_S$ and $F_0^{BR_S}(m_\pi^2)$ are constant.
This hypothesis will be assessed by the quality of the fit obtained with our model.
We shall take $R_S\equiv f_0(980)$ for the evaluation
of  $F_0^{BR_S}(m_\pi^2)$ and we use $F_0^{BR_S}(m_\pi^2)=0.13$~\cite{El-Bennich_PRD79}.

For the pion decay constant we take $f_\pi=0.1304$~GeV~\cite{PDG}. 
The $R_P$ decay constant is denoted by $f_{R_P}$ and  the $B$-meson mass by $M_B$.
Since the $\pi^+\pi^-$ $P$-wave is largely dominated by the $\rho(770)$ meson we choose $f_{R_P}=f_\rho=0.209$~GeV~\cite{bene03}.
The quantity $B_0=-2\ho\langle0\vert \bar qq\vert 0\rangle/f_\pi^2$ is proportional to the quark condensate.
We calculate it as $B_0\simeq m_\pi^2/(m_u+m_d)$.
At the renormalization scale $\mu=m_b/2$ we use $m_b=4.9$~GeV and $m_u=m_d=0.005$~GeV.
For the transition form factor between
the $B$ meson and $R_P$ state we set $A_0^{B R_P}(m_\pi^2)=0.37$~\cite{Ball_PRD58_094016}.

For the $B\pi $ scalar and vector transition form factors $F_{0}^{B\pi}(s)$ and $F_{1}^{B\pi}(s)$, we use  the following light-cone sum rule parametrization developed in Appendix A of Ref.~\cite{Ball_PRD71_014015}, viz.

\begin{eqnarray}
\label{F0pipi}
F_0^{B\pi}(s)&=& \frac{0.258}{1-s/s_0},\\
\label{F1pipi}
F_1^{B\pi}(s)&=& \frac{0.744}{1-s/M_{B^*}^2}- \frac{0.486}{1-s/s_1},
\end{eqnarray}
with $s_0=33.81$~GeV$^2$, $M_{B^*}=5.32$~GeV and $s_1=40.73$~GeV$^2$.
The pion non-strange scalar and vector form factors $\Gamma_1^{n*}(s)$ and $F_1^{\pi \pi}(s)$ will be discussed in the next section. 
Note that~\cite{meis01}
\begin{equation}
\label{gamman}
\Gamma_1^{n*}(s)=\frac{\sqrt{3}}{2B_0} \left \langle  [\pi^+{\pi^-}]_{S}\vert \bar n n\vert 0\right \rangle,
\end{equation}
with  $\bar nn=\dfrac{1}{\sqrt{2}}(\bar uu+\bar dd)$.

The transition form factor between the $B$ meson and the $R_D$ state $F^{B R_D}(m_\pi^2)$ is not 
well known ~\cite{Kim_PRD67_014002}, so it will be taken as a free parameter to be fitted.
The expressions of the tensor angular distribution factor $D(s_{12},s_{23})$ and of the $R_D$ mass 
dependence width $\Gamma(s_{23})$, similar to those used for the $f_2(1270)$ contribution 
in the BABAR Collaboration Dalitz plot analysis~\cite{Aubert:2009},
are displayed in Sec.~\ref{DwaveFBRD} of the Appendix~\ref{appendix}. The expression of the $f_2(1270)$ 
coupling to $\pi\pi$, $G_{f_2}$, is also given there.

In summary, from the $S$-, $P$- and $D$-wave matrix elements~(\ref{uv}),  (\ref{uw} and (\ref{u}),
 we obtain the total symmetrized amplitude for the $B^- \to \pi^+ \pi^- \pi^-$ decay  as
\begin{eqnarray}
\label{Mmoinsinv}
\mathcal{M}^-_{sym}(s_{12},s_{23})
=
\frac{1}{\sqrt{2}}\left [\mathcal{M}^-_S(s_{12})+
\mathcal{M}^-_S(s_{23}) + \mathcal{M}^-_P(s_{12}) (s_{13}-s_{23})
\right.
 \nonumber\\
+
\left. 
\mathcal{M}^-_P(s_{23}) (s_{13}-s_{12})
+ 
\mathcal{M}^-_D(s_{12})D(s_{23},s_{12})+\mathcal{M}^-_D(s_{23})D(s_{12},s_{23})\right ],
\end{eqnarray}
with
\begin{eqnarray}
\label{MSsij}
\mathcal{M}^-_S(s_{ij})=\frac{G_F}{\sqrt{3}}\left[-\chi_S f_\pi \left(M_{B}^2-s_{ij}\right)
F_0^{BR_S}(m_\pi^2) u(R_S \pi^-)
\right.
\nonumber\\ 
\left.
  + B_0 \frac {M_B^2-m_\pi^2}{m_b-m_d} 
F_0^{B\pi}(s_{ij}) v(\pi^-R_S)\right] \Gamma_1^{n*}(s_{ij}),
\end{eqnarray}
\begin{equation}
\label{MPsij}
\mathcal{M}^-_P(s_{ij})=\frac{G_F }{\sqrt{2}} \ho \left[
N_P\frac{f_\pi}{f_{R_P}} A_0^{B R_P}(m_\pi^2) u(R_P \pi^-)+
F_1^{B\pi}(s_{ij})w(\pi^- R_P) \right] F_1^{\pi \pi}(s_{ij}),
\end{equation}
and

\begin{equation}
\label{MDsij}
\mathcal{M}^-_D(s_{ij})=-\frac{G_F }{\sqrt{3}} u(R_D \pi^-) \ho  \frac{f_\pi}{\sqrt{2}}
   F^{B R_D}(m_\pi^2) \ho \frac{G_{f_2}}{m^2_{R_D}-s_{ij}-im_{R_D}\Gamma(s_{ij})}.
\end{equation}
For the fully symmetrized $B^+\to\pi^+\pi^- \pi^+$ decay amplitude we have

\begin{eqnarray}
\label{Mplusinv}
\mathcal{M}^+_{sym}(s_{12},s_{23})
=
\frac{1}{\sqrt{2}}\left [\mathcal{M}^+_S(s_{12})+
\mathcal{M}^+_S(s_{23}) + \mathcal{M}^+_P(s_{12}) (s_{13}-s_{23})
\right. \nonumber\\
+
\left.
\mathcal{M}^+_P(s_{23})(s_{13}-s_{12})
+ 
\mathcal{M}^+_D(s_{12})D(s_{23},s_{12})+\mathcal{M}^+_D(s_{23})D(s_{12},s_{23})\right],
\end{eqnarray}
with

\begin{equation}
\label{M+sij}
\mathcal{M}^+_{S,P,D}(s_{ij})=\mathcal{M}^-_{S,P,D}\left(s_{ij}, \lambda_u \to \lambda^*_u, \lambda_c \to \lambda^*_c, B^-\to B^+\right).
\end{equation}

\section{Scalar and vector form factors}
\label{SVff}
As shown in Ref.~\cite{Barton65} the full knowledge of strong interaction meson-meson form factors  is available if the meson-meson interaction is known at all energies.
The calculation of the $S$- and $P$-wave amplitudes (\ref{MSsij}) and (\ref{MPsij}) requires the values of the scalar and vector $B\pi$, $B(\pi\pi)$ and pion form factors.
The knowledge of the $B\ \to \pi$ and $B\to [\pi \pi]_{S,P}$ transition form factors is needed far below the $B\pi $ and $B [\pi \pi]_{S,P}$ scattering region.
 One has then to rely on theoretical models constrained by experiment, as we do here for the $B[\pi \pi]_{S}$  form factor, using the value (see above in the previous section)  determined in Ref.~\cite{El-Bennich_PRD79}.
One could also use covariant light-front model, like that of Ref.~\cite{Cheng_PRD69} or, if available, semi-leptonic decay analysis results.
For the $B\pi$ form factors we take the QCD light-cone sum rule results of Ref.~\cite{Ball_PRD71_014015} recalled above in Eqs.~(\ref{F0pipi}) and (\ref{F1pipi}).
The special case of the pion form factors is developed below.

\subsection{\bf{The pion scalar form factor}}
\label{pipiSff}
In the $\pi \pi$ case, the low-energy $S$ wave being known and modeling the high-energy part  one can rely on the Muskhelishvili-Omn\`es  equations~\cite{MO} to build up the pion scalar form factors.
Their evaluation from these equations has been discussed in Ref.~\cite{DonoghueNPB343} and followed and developed in Ref.~\cite{L6}.
However here, we shall use another approach, initiated in Ref.~\cite{meis01} and applied, using a different $\pi \pi$ scattering matrix, in Ref.~\cite{fkll}.
Extending this last work by introducing three channels and keeping the off-shell contributions, the pion scalar form factor $\Gamma_1^{n*}(s)$ entering in the $S$-wave amplitude Eq.~(\ref{MSsij}) is modeled according to the following relativistic three coupled-channel equations

\begin{equation}
\label{gammai}
\Gamma_i^{n*}(s)=R_i^{n}(E)+\sum_{j=1}^3 R_j^{n}(E) H_{ij}(E),\quad i=1,2,3,
\end{equation}
with

\begin{equation}
\label{Hij}
H_{ij}(E)=\int \frac{d^3p}{(2\pi)^3} T_{ij}(E,k_i,p) \frac{1}{E-2\sqrt{p^2+m_j^2}+i\epsilon} \ \frac{k_j^2+\kappa^2}{p^2+\kappa^2},
\end{equation}
where $E$ represents the total energy, i.e., in the $\pi \pi$ center of mass, $E=\sqrt{s}$ and $p$ is the off-shell momentum. 
In Eqs~(\ref{gammai}) and (\ref{Hij}), the indices $i, j=1, 2, 3$ refer to the $\pi \pi$, $K \bar K$ and effective $(2\pi)(2\pi)$  channels, respectively. 
The center of mass momenta are $k_j=\sqrt{s/4-m_j^2}$, with $m_1=m_\pi$, $m_2=m_K$ and $m_3=m_{(2\pi)}$.  
The  $T$ matrix is the  corresponding three-channel two-body scattering matrix.
Here we use the solution $A$ of the three-coupled channel model of Refs.~\cite{Kaminski:1997gc, Kaminski_EPJC9_141}, where the effective $m_{(2\pi)}$= 700 MeV. 
The functions  $R_i^{n}(E)$ are the production functions responsible for the formation of the meson pairs before their scattering.
From Eqs.~(\ref{gammai}) and (\ref{Hij}) one can check that

\begin{equation}
\label{Uni}
Im\ \Gamma_i^{n*}(s)=- \sum_{j=1}^3 \frac{k_j\sqrt{s}}{8\pi}T_{ji}^*(E,k_j,k_i) \Gamma_j^{n*}(s)\theta (\sqrt{s}-2m_j).
\end{equation}
This is the same unitary relation  as that of the corresponding Muskhelishvili-Omn\`es pion scalar form factors constructed in Ref.~\cite{L6} [see Eq.~(28) therein].

In Eq.~(\ref{Hij}) the regulator function $(k_j^2+\kappa^2)/(p^2+\kappa^2)$, which reduces to 1 on-shell ($k_j=p$), ensures the convergence of the integral. 
The range parameter $\kappa$ will be fitted to data.
The choice of a separable form for the interaction yields analytic expressions for the $T$ matrix elements.
One introduces a rank-2 separable potential in the $\pi \pi$ channel and a rank-1 separable potential in the $K\bar K$ and in the $(2\pi)(2\pi)$ ones.
According to the formalism developed in Ref.~\cite{KLM_PRD50_3145} and applied in Ref.~\cite{Kaminski:1997gc} one has for the  $T$ matrix elements:

\begin{eqnarray}
\label{T1j}
T_{11}(E,p,k_1) &=&g_0(k_1)t_{00}(E)g_0(p)+g_1(k_1)t_{11}(E)g_1(p)+g_0(k_1)t_{10}(E)g_1(p)\nonumber \\
&+&g_1(k_1)t_{01}(E)g_0(p), \nonumber \\
T_{21}(E,p,k_1)&=& g_0(k_1)t_{02}(E)g_2(p)+g_1(k_1)t_{12}(E)g_2(p), \nonumber \\
T_{31}(E,p,k_1)&=& g_0(k_1)t_{03}(E)g_3(p)+g_1(k_1)t_{13}(E)g_3(p),
\end{eqnarray}
where

\begin{eqnarray}
\label{g0jR}
 g_0(k_1)&=&\sqrt{\frac{4\pi}{m_{\pi}}}\frac{1}{k_1^2+\beta_0^2}, \nonumber \\
  g_j(k_i)&=&\sqrt{\frac{4\pi}{m_i}}\frac{1}{k_i^2+\beta_j^2}, \quad j=1,2,3.
\end{eqnarray}
The parameters $\beta_j$, $j=0, 1, 2, 3$,  of the separable form  of the scattering $T$ matrix are given in Table~1 of Ref.~\cite{Kaminski:1997gc} (fit $A$).

One can extend the expressions of the reduced symmetric $t(E)$ matrix elements given in terms of the separable potential parameters  in Appendix A of Ref.~\cite{KLM_PRD50_3145} to the case of Ref.~\cite{Kaminski:1997gc} which we use here.
The Yamaguchi form~\cite{Yamaguchi} of the $g_0(p)$ and $g_i(p)$~(\ref{g0jR}) in the
 $T$ matrix elements (\ref{T1j}) leads the following analytic expression for
  $\Gamma_i^{n*}(s)$ in Eq.~(\ref{gammai})

\begin{equation}
\label{G1n$}
\begin{split}
\Gamma_1^{n*}(s)=&R_1^n(E)+R_1^n(E)\{[t_{00}(E)g_0(k_1)+t_{01}(E)g_1(k_1)]g_0(k_1)F_{10}(k_1)+\\
&[t_{11}(E)g_1(k_1)+t_{10}(E)g_0(k_1)]g_1(k_1)F_{11}(k_1)\}+\\
&R_2^n(E)[g_0(k_1)t_{02}(E)+g_1(k_1)t_{12}(E)]g_2(k_2)F_{22}(k_2)+\\
&R_3^n(E)[g_0(k_1)t_{03}(E)+g_1(k_1)t_{13}(E)]g_3(k_3)F_{33}(k_3),
\end{split}
\end{equation}
where

\begin{eqnarray}
\label{Fij}
 F_{10}(k_1)&=&\frac{I_{1,0}(k_1)}{g_0(k_1)h_0(k_1)}, \nonumber \\
 F_{11}(k_1)&=&\frac{I_{1,1}(k_1)}{g_1(k_1)h_1(k_1)}, \nonumber \\
 F_{22}(k_2)&=&\frac{I_{2,2}(k_2)}{g_2(k_2)h_2(k_2)}, \nonumber  \\
 F_{33}(k_3)&=&\frac{I_{3,3}(k_3)}{g_3(k_3)h_3(k_3)}, 
\end{eqnarray}
with

\begin{eqnarray}
\label{hi}
 h_i(k_i)&=&\sqrt{\frac{4\pi}{m_i}}\frac{1}{k_i^2+\kappa^2}, \quad i=1,2,3,\nonumber \\
  h_0(k_1)&=& h_1(k_1),
\end{eqnarray}
and

\begin{equation}
 I_{i,j}(k_i)=\int\frac{d^3p}{(2\pi)^3} \ho g_j(p)\ho \frac{1}{E-2\sqrt{p^2+m_i^2}+i\epsilon}h_{i}(p),
\end{equation}
where $E=2 \sqrt{k_i^2+m_i^2}$, $i=1,2,3$. 
The analytical expression for these integrals can be found in Appendix $A$ of Ref.~\cite{KLM_PRD50_3145}.

As in Ref.~\cite {meis01} one constraints the $\Gamma_i^{n*}(s)$ to satisfy the low energy behavior given by next-to-leading order one loop calculation in chiral perturbation  theory (ChPT). 
One  writes the expansion at low $s$ as
\begin{equation}
\label{GlowE}
\Gamma_i^{n}(s) \cong d_i^{n}+f_i^{n}s, \quad i=1,2,3,
\end{equation}
with real coefficients, $\Gamma_i^{n}(s)$ being real below the $\pi\pi$ threshold.
Using the expressions  obtained in NLO in ChPT for the $\Gamma_i^{n*}(s)$  given in
Refs.~\cite{meis01, LM_PRD74_034021} one gets,

\begin{eqnarray}
\label{df1n}
 d_1^{n}&=&\sqrt{\frac{3}{2}} \left [ 1+\frac{16m_\pi^2}{f^2}\left( 2L_8^r-L_5^r\right ) +8 \frac{2m_K^2+3m_\pi^2}{f^2}\left( 2L_6^r-L_4^r\right )\right.\nonumber \\
 &&+  \left. \frac{m_\pi^2}{36\pi^2f^2} +\frac{m_\pi^2}{16\pi^2f^2} \log{\frac{m_\pi^2}{\nu^2}}
 -  \frac{1}{96\pi^2f^2}\left(\frac{m_\pi^2}{3}+m_\eta^2\right)\log{\frac{m_\eta^2}{\nu^2}}\right],\nonumber \\
  f_1^{n}&=&\sqrt{\frac{3}{2}} \left [ \frac{4}{f^2}\left( 2L_4^r+L_5^r\right ) - \frac{1}{16\pi^2 f^2}\left(1+\log{\frac{m_\pi^2}{\nu^2}}\right)
\right.\nonumber \\
 &&-  \left. \frac{1}{64 \pi^2 f^2} \left(1+\log{\frac{m_K^2}{\nu^2}}\right)
 -  \frac{m_\pi^2}{192 \pi^2 f^2} \left(\frac{1}{m_\pi^2}-\frac{1}{9m_\eta^2}\right)\right],
\end{eqnarray}
and
\begin{eqnarray}
\label{df2n}
 d_2^{n}&=&\frac {1}{\sqrt{2}} \left [ 1+ \frac{m_\eta^2}{48\pi^2f^2}\log{\frac{m_\eta^2}{\nu^2}}+\frac{16m_K^2}{f^2}\left( 2L_8^r-L_5^r\right ) 
\right.\nonumber \\
&&\left.
+8 \frac{6m_K^2+m_\pi^2}{f^2} \left( 2L_6^r-L_4^r\right )
+  \frac{m_K^2}{72\pi^2f^2}\left(1+\log{\frac{m_\eta^2}{\nu^2}}\right)\right],\nonumber \\
  f_2^{n}&=&\frac{1}{\sqrt{2}} \left [ \frac{4}{f^2}\left( 2L_4^r+L_5^r\right ) - \frac{1}{64\pi^2f^2} \left(1+\log{\frac{m_\eta^2}{\nu^2}}\right)-\frac{m_K^2}{432\pi^2f^2}\frac{1}{m_\eta^2}
\right.\nonumber \\
 &&-  \left. \frac{3}{64\pi^2f^2} \left(1+\log{\frac{m_K^2}{\nu^2}}\right)
 -  \frac{3}{64\pi^2f^2} \left(1+\log{\frac{m_\pi^2}{\nu^2}}\right)\right],
\end{eqnarray}
$\nu$ being the scale of dimensional regularization and $f=f_\pi/\sqrt{2}$ .
Furthermore for  the ChPT low-energy constants, $L_k^r$, $k=4,5,6,8$, 
we use the recent  determinations of lattice QCD at $\nu=1$~GeV as given in Table X 
of Ref.~\cite{Alton_PRD78_114509}.
For $f=92.4$~MeV, we obtain $d_1^n=1.1957$, $f_1^n=3.1329$~GeV$^{-2}$,
 $d_2^n=0.7193$ and $f_2^n=1.6719$~GeV$^{-2}$.
As in Ref.~\cite{L6}  we assume $\Gamma_3^n(0)=0$ which leads to $d_3^{n} =0$ and we also assume $f_3^{n} =0$.

The real production functions are parametrized as
\begin{equation}
\label{RjE}
 R_i^{n}(E)=\frac{\alpha_i^n+\tau_i^nE+\omega_i^nE^2}{1+cE^4},\ i=1,2,3,
\end{equation}
the fitted parameter $c$ controling the high energy behavior.
The other parameters, $\alpha_i^n$, $\tau_i^n$ and $\omega_i^n$ are calculated by requiring that $ \Gamma_i^{n}(s)$ in Eq.~(\ref{gammai}) has the low energy expansion Eq.~(\ref{GlowE}). 
These nine parameters satisfy a linear system of nine equations displayed in Appendix~\ref{alpha_i_equation}. 
Their numerical values, depending on the value of the range parameter $\kappa$ [see Eq.~(\ref{Hij})], will be given in Sec.~\ref{results}.

\subsection{\bf{ The pion vector form factor}}
\label{pipiPff}
As for the scalar case one could use the Muskhelishvili-Omn\`es equations to built up the pion vector form factor.
This was done in Ref.~\cite{Bppk} for the $K\pi$ vector form factor.
Here, noting that the knowledge  of this form factor is required to describe the $\tau^-\to\pi^-\pi^0\nu_\tau$ decay, we shall use the phenomenological model of the Belle Collaboration~\cite{Fujikawa_PRD78_072006}.
Fitting their high statistics data, they built the pion vector form factor $F_1^{\pi \pi} (s_{23})$ by including the contribution of the three vector resonances $\rho(770)$, $\rho(1450)$ and $\rho(1700)$.
Here we use the parameters given in the third column of Table VII of Ref.~\cite{Fujikawa_PRD78_072006}.

\section{Observables and data fitting}
\label{fit}

\subsection{\bf {Physical observables}}
\label{observables}

The symmetrized $B^{-} \to \pi_1^- \pi_2^+ \pi_3^-$ amplitude (\ref{Mmoinsinv}) depends on the two effective $\pi\pi$  masses, $m_{12}=\sqrt{s_{12}}$ and  $m_{23}=\sqrt{s_{23}}$ of the Dalitz plot.
In the center of mass of  $\pi^-(p_1)$  and $\pi^+(p_2)$, the pion momenta fulfill the equations  

\begin{eqnarray}
\label{kincms12}
\vert \overrightarrow{p_1} \vert  & = & \frac{1}{2}\sqrt{m_{12}^2-4m_\pi^2},\hspace{2cm} \vert \overrightarrow{p_2} \vert = \vert \overrightarrow{p_1} \vert,  \nonumber \\
\vert \overrightarrow{p_3} \vert  & = &  \frac{1}{2m_{12}}\sqrt{\left[M_B^2-\left(m_{12}+m_\pi\right)^2\right]
\left[M_B^2-\left(m_{12}-m_\pi\right)^2\right]}, 
\end{eqnarray}
and the cosine of the helicity angle $\theta$ between the direction of $
\overrightarrow{p_2}$ and that of $\overrightarrow{p_3}$ reads

\begin{equation}
\label{costeta23}
\cos \theta =\frac{1}{2\vert \overrightarrow{p_2} \vert  \vert \overrightarrow{p_3} \vert} 
\left[-m_{23}^2+\frac{1}{2}\left(M_B^2-m_{12}^2+3m_\pi^2\right)\right].
\end{equation}
For fixed values of the effective mass $m_{12}$, the variables $\cos \theta$ and $m_{23}$ are equivalent. 

The double differential $B^- \to \pi^- \pi^+ \pi^-$ branching fraction is

\begin{equation}
\label{ddbr}
\frac{d^2\mathcal{B}^-}{dm_{12}\ d\cos \theta}=\frac{1}{\Gamma_B}\frac{m_{12}\vert \overrightarrow{p_2} \vert \vert \overrightarrow{p_3} \vert }{8 (2\pi)^3 M_B^3} \left \vert \mathcal{M}^-_{sym}(s_{12},s_{23})\right \vert^2,
\end{equation}
where $\Gamma_B$ is the total width of the $B^-$. 
Since the Dalitz plot is symmetric under the interchange of $m_{12}$ and  $m_{23}$, one can limit the integration range on $m_{23}$ to the values larger than $m_{12}$; hence,  the  differential effective mass distribution reads 
\begin{equation}
\label{dBdm12}
\frac{d\mathcal{B}^-}{dm_{12}}=\int_{-1}^{\cos \theta_g} \frac{d^2\mathcal{B}^-}{dm_{12}\ d\cos \theta} \ho d\cos \theta,
\end{equation}
where $\cos \theta_g$ corresponds to the value of $\cos \theta$ in Eq.~(\ref{costeta23}) with $m_{12}=m_{23}$, viz.,

\begin{equation}
\label{costhetag}
\cos \theta_g =\frac{1}{4\vert \overrightarrow{p_2} \vert  \vert \overrightarrow{p_3} \vert}  \left(M_B^2-3m_{12}^2+3m_\pi^2\right).
\end{equation}
The variable $m_{12}$ in Eq.~(\ref{dBdm12}) is also called the light  (or minimal) effective mass $m_{min}$ while $m_{23}$ is the heavy (or maximal) effective mass, $m_{max}$.
The  $B^{-} \to \pi^- \pi^+ \pi^-$ branching fraction is then twice the integral of the differential branching fraction (\ref{dBdm12}) over $m_{12}$.

\begin{figure}

 \includegraphics[angle=0,width=0.9\columnwidth]{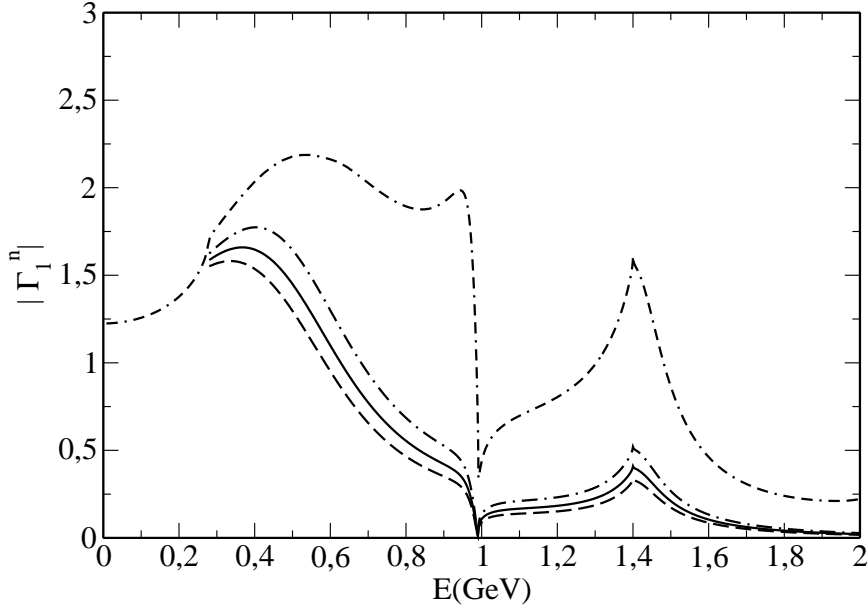}~~~~

\caption{\label{Gamma1N} Modulus of the pion scalar form factor  $\Gamma_1^n$ (solid line), obtained in 
our fit
 using the NLO $a_i^p$ with $\kappa=2$~GeV and for which the fitted parameter  $c=(19.5 \pm 4.2)$~GeV$^{-4}$, 
compared to that  calculated in Ref.~\cite{PrivateBM} using the Muskhelishvili-Omn\`es equations (double-dash dot line). 
The dash-dot line (for $c=15.3$~GeV$^{-4}$)
and the dashed one (for $c=23.7$~GeV$^{-4}$) represent the variation of the  $\Gamma_1^n$ modulus when 
$c$ varies within its error band.}
\end{figure}

\subsection{\bf{Data fitting}}
\label{Observables and data fitting}

We aim at  describing the experimental $\pi^+\pi^-$ distributions obtained by the BABAR
Collaboration in the Dalitz plot analysis of the $B^{\pm} \to \pi^{\pm}\pi^{\pm}\pi^{\mp}$ decays~\cite{Aubert:2009}. 
Two different background  distributions, related to the $q\bar{q}$ and the $B\bar{B}$ components, are subtracted from Fig.~4 of Ref.~\cite{Aubert:2009}. 
Six light effective $\pi^+\pi^-$ mass distributions are extracted for $B^+$ and $B^-$ decays with a subdivision of the data into positive and negative values of the cosine of the helicity angle $\theta$.
For the  $B^+$ and $B^-$ distributions we reject two data points corresponding to the $\pi^+\pi^-$ effective masses equal to 485 and 515 MeV. 
Also two points at 470 and 530 MeV for the four mass distributions with $\cos\theta >0$ or with $\cos\theta <0$ are not taken into account.
This is done to exclude the possible contribution of the decay processes $B^{\pm} \rightarrow K^0_S \pi^{\pm}$. 

As a by-product of the background subtraction, five data points, with a small number of events, have negative values with small statistical errors. 
For these five  data points  we  increase their errors
to values corresponding to those of the points lying in a close vicinity.
This is done at 1385 MeV for the $B^-$ 
distribution, at 1475 MeV for the $B^+$ one, at 290 and 1610 MeV for the $B^-$ distribution with $\cos\theta >0$ and at 1490 MeV  for the $B^-$ one with $\cos\theta <0$.

We perform a $\chi^2$ fit to the 170 data points corresponding to the six invariant mass 
distributions described above. In addition, we include the experimental branching ratio for the 
$B^\pm \to \rho(770)^0 \pi^\pm,\  \rho(770)^0 \to \pi^+ \pi^-$ decay channel.
The theoretical distributions are normalized to the number of experimental events in the analyzed range from 290 up to 1640 MeV.
In the fits, done for a fixed value of the range parameter $\kappa$ entering  Eqs.~(\ref{Hij})
[see Sec. 5], the following four parameters were varied: 
the production functions $R_i^n(E)$ [Eq.~(\ref{RjE})] parameter $c$, the real $S$-wave strength 
parameter $\chi_S$, the real $P$-wave correction parameter $N_P$ [Eq.~(\ref{XP})] and the transition form 
factor $F^{B R_D}(m_\pi^2)$ [Eq.~(\ref{XD})]. 
 
\begin{figure}

  \includegraphics[height=.35\textheight]{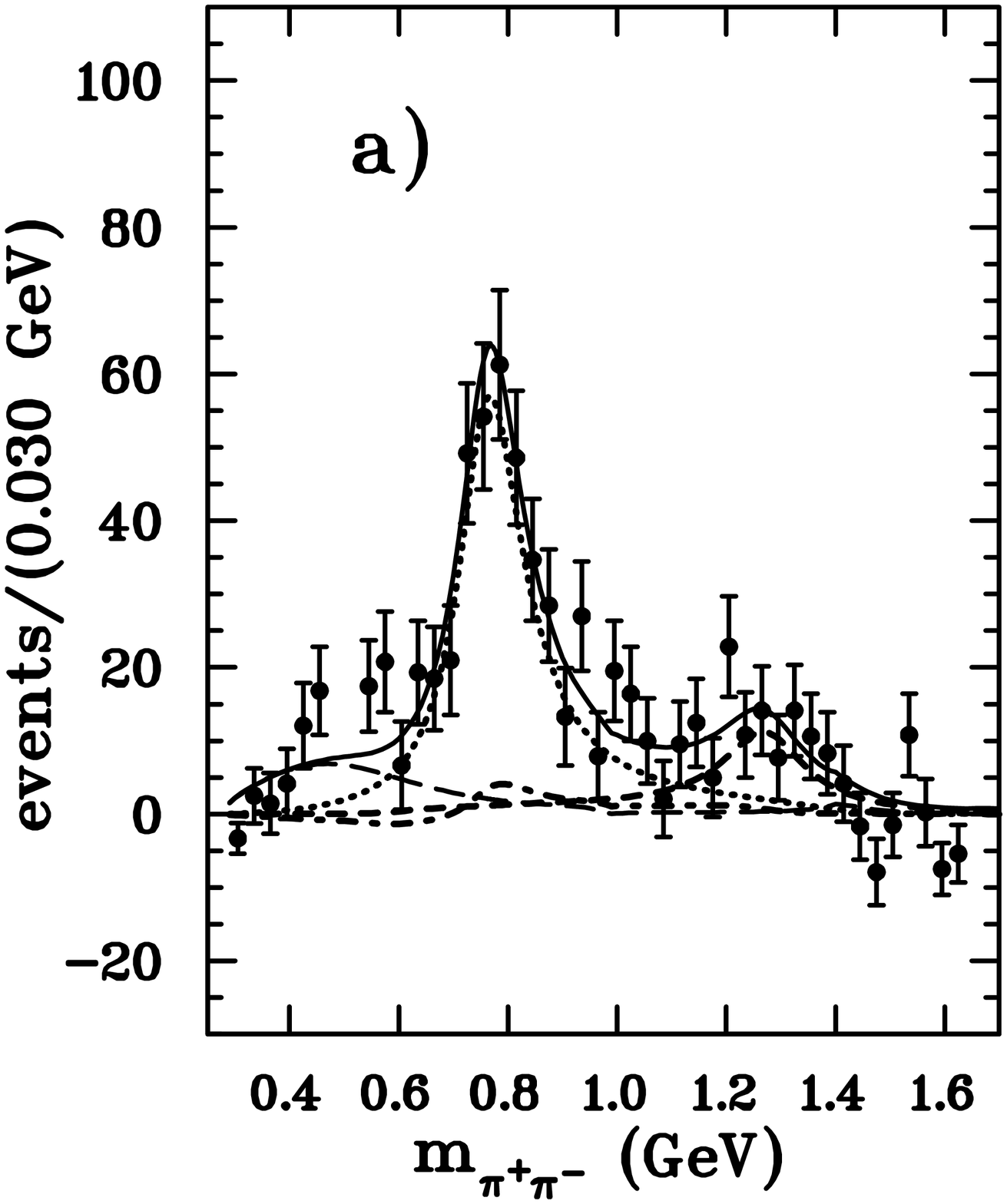}~~~~
   \includegraphics[height=.35\textheight]{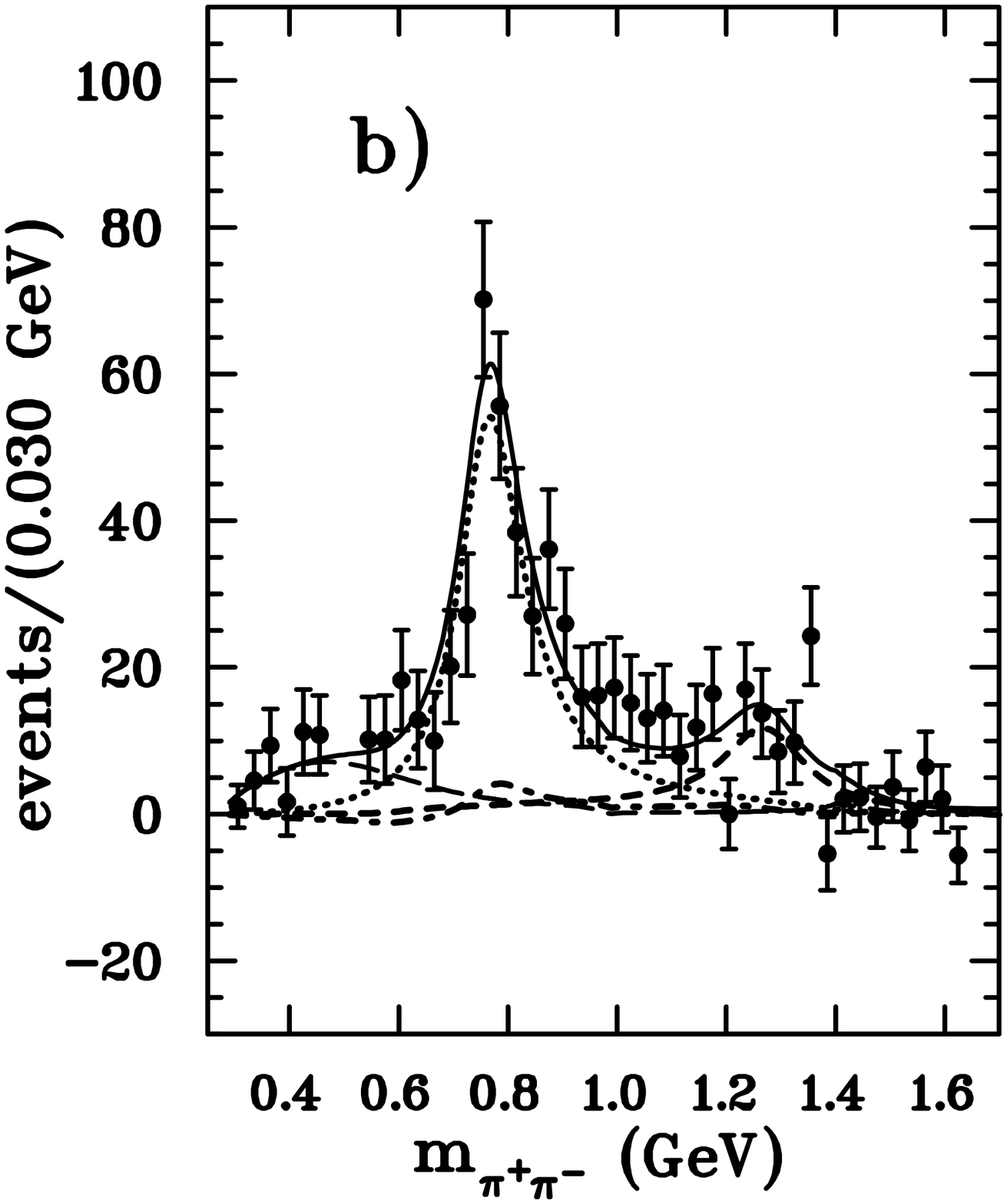}~~

\vspace*{6pt}

\caption{\label{lightB-+} The $\pi^+\pi^-$ light effective mass distributions from the fit to the BABAR experimental data~\cite{Aubert:2009}, a) for the $B^-$ decays and b) for the $B^+$ decays. 
The long-dash line represents the $S$-wave contribution of our model, the dot line 
that of the $P$ wave, the short-dash line that of the $D$ wave and the dot-dash line that of the interference term. 
The solid line corresponds to the sum of these contributions.}
\end{figure}

\begin{figure}
  \includegraphics[height=.35\textheight]{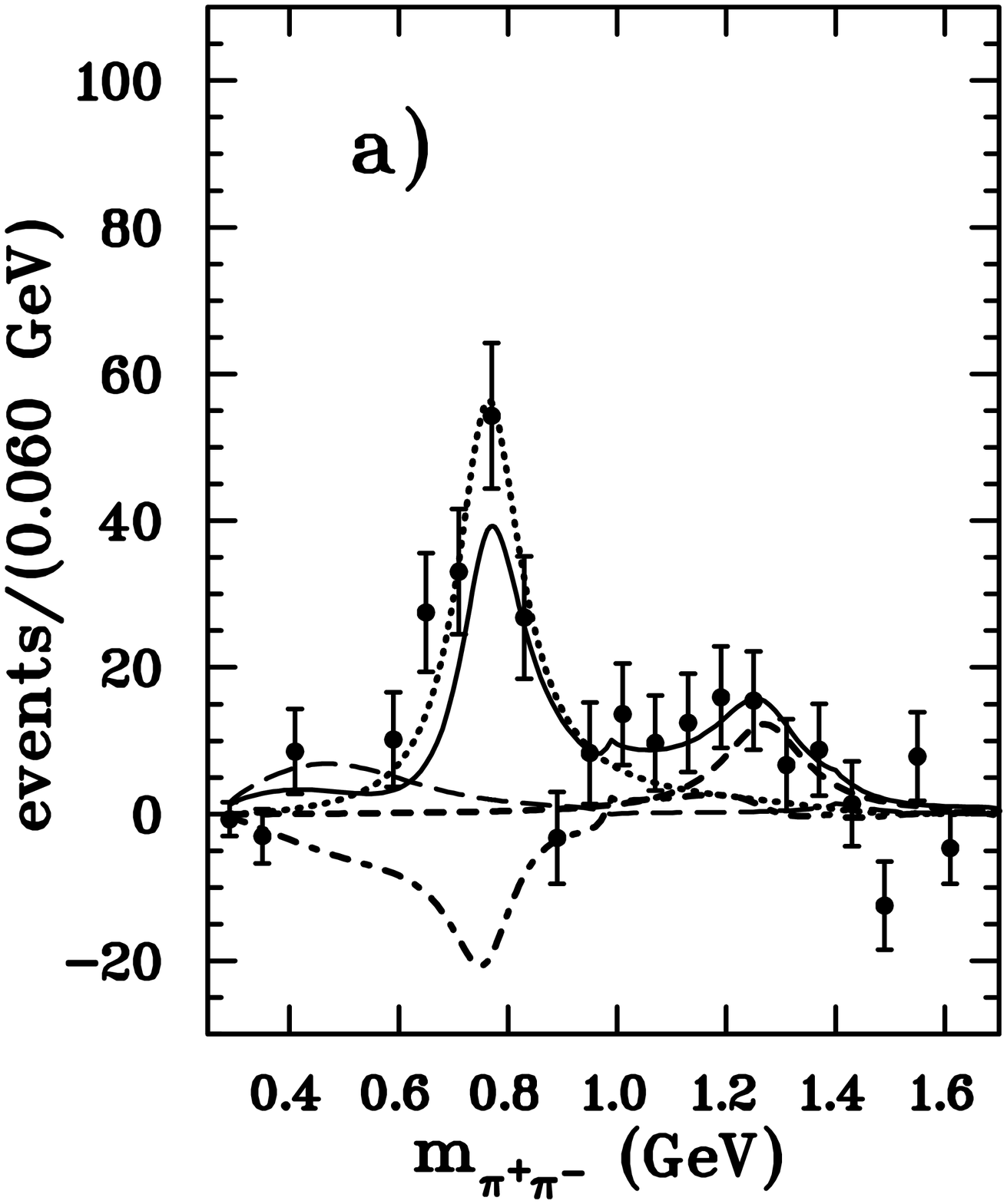}~~~~
   \includegraphics[height=.35\textheight]{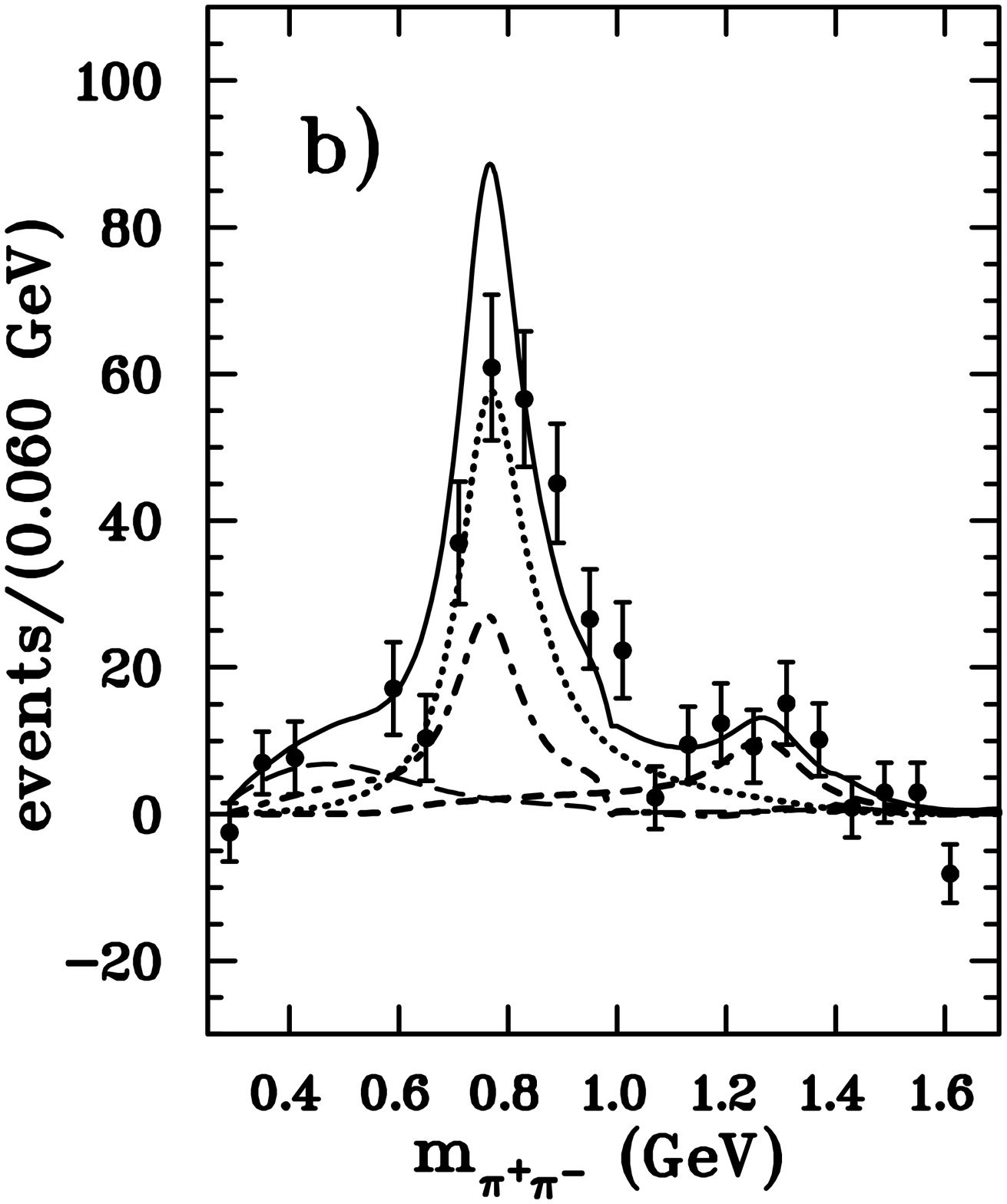}~~

\vspace*{6pt}

\caption{\label{lightB-cos-+} As in Fig.~\ref{lightB-+} but for the $B^-$ decays a) with $\cos \theta <0$  and b) with $\cos \theta >0$.}

\end{figure}

\begin{figure}
 
  \includegraphics[height=.35\textheight]{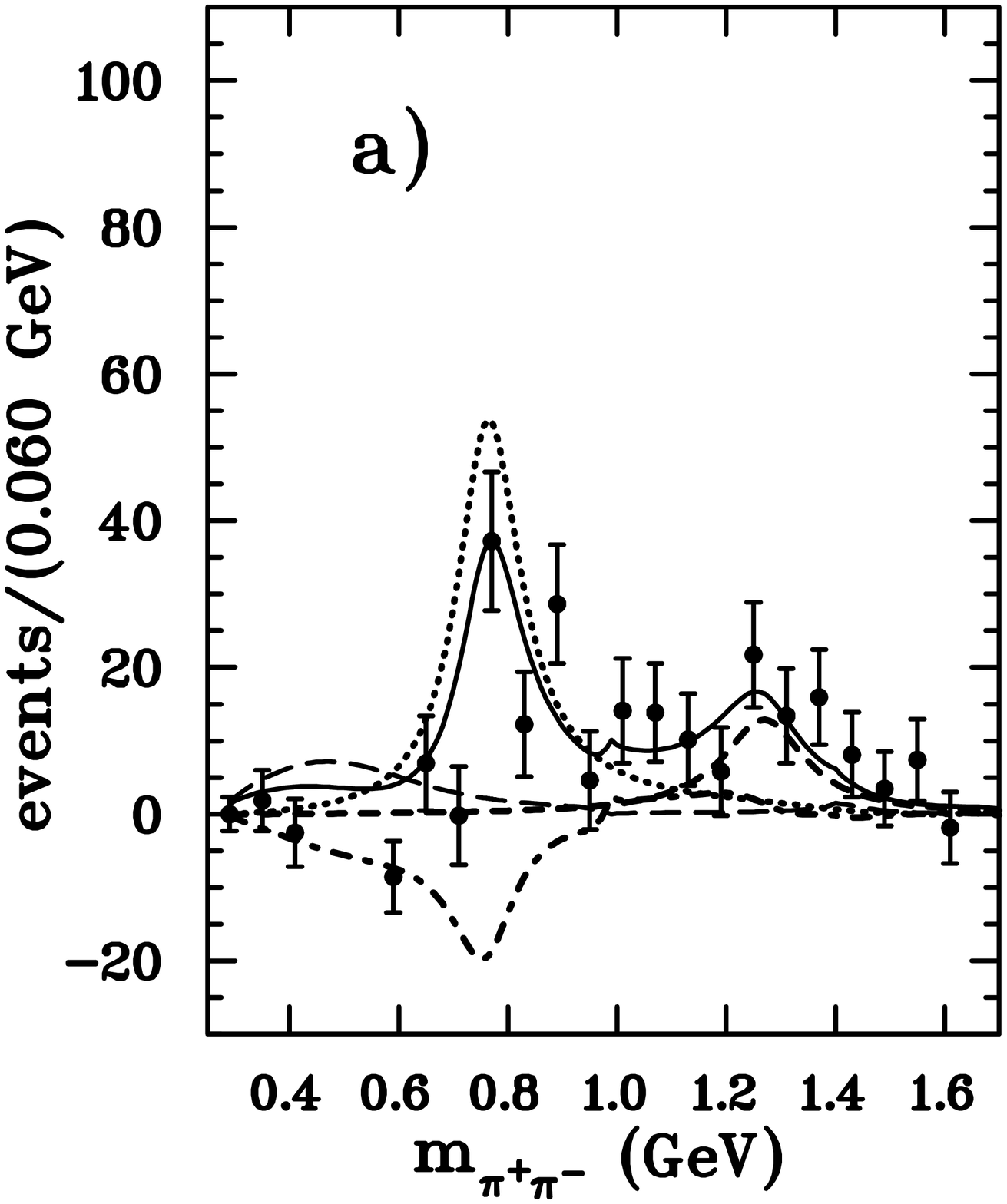}~~~~
   \includegraphics[height=.35\textheight]{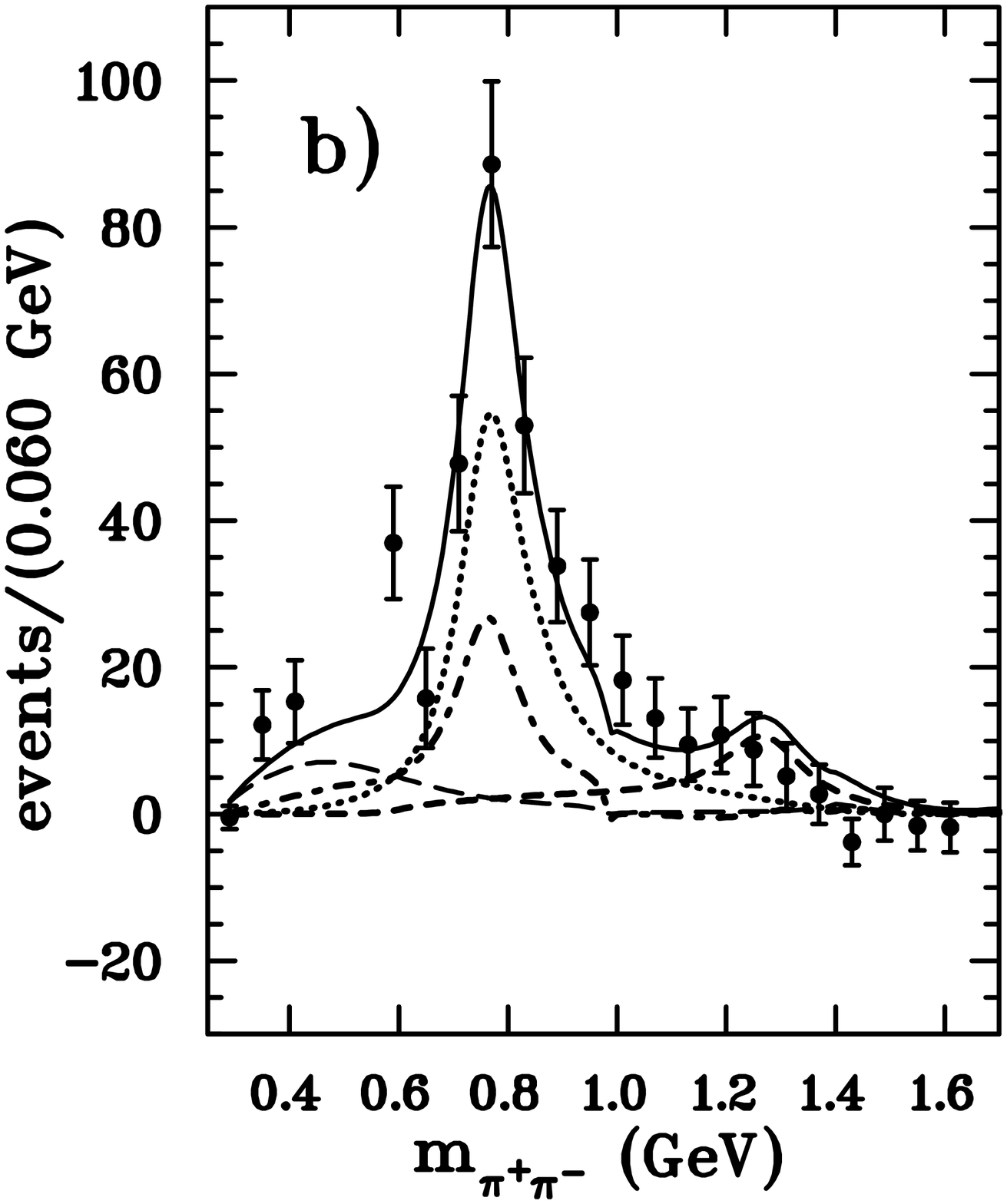}~~

\vspace*{6pt}

\caption{\label{lightB+cos-+} As in Fig.~\ref{lightB-cos-+} but  for the $B^+$ decays.}
\end{figure}

\section{Results and discussion}
\label{results}
In the fits to the selected BABAR data as described in the previous section, the CKM matrix elements [see Eq.~(\ref{lambda})] are calculated with $\lambda= 0.2257$, $A=0.814$, $\bar \rho=0.135$ and 
$\bar \eta=0.349$~\cite{PDG} which leads to
 $\lambda_u=1.26\times 10^{-3}-i\  3.27\times 10^{-3}$ and
 $\lambda_c=-9.35\times 10^{-3}-i\ 1.72\times 10^{-6}$.
The LO contributions of the Wilson coefficients to  the $a_i^p$ Eq.~(\ref{aip})
are given in the second and fourth columns of Table~\ref{Tabailovp}.
The sum of the leading order coefficient plus the next-to-leading order vertex and penguin corrections for
 the  $a_i^p$ coefficients, entering into $u(R_{S,P} \pi^-)$ [Eq.~(\ref{RSpi})], $v(\pi^- R_S)$
[Eq.~(\ref{piRS})] and $w(\pi^- R_P)$ [Eq.~(\ref{wRP})], are displayed in columns three and five,
 respectively. It can be seen that the NLO corrections are relatively small except for the coefficient
 $a_2$ which,
however, has only a small contribution to the decay amplitude.
The corrections are calculated according to Refs.~\cite{Beneke:2001ev} and ~\cite{bene03} using the Gegenbauer moments for pions taken from the Table~2 of Ref.~\cite{Beneke:2001ev} and  the corresponding moments for the $\rho$ meson from Table~1 of Ref.~\cite{BallJones2007}. 
In the calculation of the coefficients $a_6^p(\pi^- R_S)$ and  $a_8^p(\pi^- R_S)$, contributing to $v(\pi^- R_S)$, we apply the method explained in Appendix A of Ref.~\cite{Cheng:2005nb}.
Here the renormalization scale $\mu=m_b/2$ and we take for the strong coupling constant $\alpha_s(m_b/2)= 0.303$.

\begin{table}
\caption{
\label{Tabailovp} 
Leading order (LO) and next-to-leading order (NLO) coefficients  $a_i^p(R_{S,P}\pi^-)$, $a_i^p( \pi^-R_{S})$ (in parentheses) and  $a_i^p( \pi^-R_{P})$ [see Eq.~(\ref{aip})] entering into $u(R_{S,P} \pi^-)$ [Eq.~(\ref{RSpi})], $v(\pi^- R_S)$[Eq.~(\ref{piRS})] and $w(\pi^- R_P)$ [Eq.~(\ref{wRP})], respectively.
The NLO coefficients are the sum of the LO coefficients plus next-to-leading order vertex and penguin corrections.
Here the renormalization scale is $\mu=m_b/2$.
The superscript $p$  is omitted for $ i=1$, $2$, $3$, $5$, $7$ and $9$, the penguin corrections being zero for these cases.
}
\begin{tabular}{ccccccc}
            &\multicolumn{2}{c}{$a_i^p(R_{S,P}\pi^-)$~~~~~~~~~}&\multicolumn{2}{c}{$a_i^p(\pi^-R_{S,P})$~~~~~~~~~ }\\
            &       LO     &    NLO                            &      LO    &  NLO                 \\
\hline
$a_1$       & $1.039$      & $1.071  + i 0.03$                 &            &                       \\
$a_2$       &              &                                   &   $ 0.084$ &  $-0.041 -i 0.114$    \\
$a_4^{u}$   &$    -0.044 $ & $-0.032  - i 0.019$               &  $-0.044$  &  $-0.032  - i 0.019$  \\
$a_4^c$     & $ -0.044 $   & $ -0.039  - i 0.007$              &  $-0.044$  &  $ -0.039  - i 0.007$\\
$a_6^{u}$   & $-0.062 $    & $ -0.057  - i 0.017  $            & $(-0.062)$ & $(-0.075-i 0.017$)   \\
$a_6^c$     &$  -0.062 $   & $ -0.062  - i 0.004$              & $(-0.062)$ & $(-0.079-i 0.004$)   \\
$a_7$       &              &                                   &  $0.0001$  &  $0.0 +i 0.0001$   \\
$a_8^{u}$   &$0.0007$      & $ 0.0008 + i 0.0$                 & $(0.0007)$ & $(0.0007+i 0.0)$    \\
$a_8^c$     &$0.0007$      & $0.0008 + i 0.0 $                 & $(0.0007)$ & $(0.0007+i 0.0)$       \\
$a_9$       &              &                                   &  $-0.0094$ &  $-0.0097 -i 0.0003$    \\
$a_{10}^{u}$& $-0.0009 $   & $0.0006 + i 0.0010$               &  $-0.0009$ &   $0.0006 + i 0.0010$  \\
$a_{10}^c$  &  $-0.0009$   & $0.0006 + i 0.0010$               &  $-0.0009$ &   $0.0006 + i 0.0010$
\end{tabular}

\end{table}
\begin{table}
\caption{\label{Tabalphai}Parameters of the production functions $R_i^n(E)$ Eq.~(\ref{RjE}) for  $\kappa$ = 2 GeV}
\begin{tabular}{cccc}
      i & $\alpha_i^n$ & $\tau_i^n$ (GeV$^{-1})$ & $\omega_i^n$ (GeV$^{-2})$\\
\hline 
     1 &  $ 0.7095 $   & $-0.2707$                        & $ 1.6251$\\
 
     2 &  $0.5759$     & $-0.0032$                         & $1.4171$  \\ 
 
     3 &  $1.003$     & $0.3724$                        & $2.7427$  \\ 
\end{tabular}
\end{table}
There are five free parameters at our disposal.  
Two of them, the regulator range $\kappa$ and the high energy cut-off $c$ of the production functions 
[Eq.~(\ref{RjE})] are linked to the determination of the $S$-wave $\Gamma_1^n$ form factor.
The other three, $\chi_S$, $N_P$ and $F^{B R_D}(m_\pi^2)$ are related to the strength of the $S$, $P$ and
$D$ amplitudes, respectively.
The range $\kappa$ should be larger than 0.8 GeV which is the on-shell pion 
momentum approximately equal to the half of the effective $m_{\pi \pi}$ upper limit $\sim 1.64$~GeV which we
used.
In our fits we find that the total $\chi^2$ decreases slowly when $\kappa$ decreases from the high 
value of 5~GeV.
Here we fix the range parameter $\kappa$ to be 2~GeV.
We perform two fits for the full $S+P+D$-wave amplitude calculated with the NLO and with the LO $a_i^p$ 
coefficients.
Hereafter the quoted results given inside parentheses correspond to the numbers obtained in the second 
fit. 
The quoted errors on our results come from the statistical errors in the experimental data.

A good overall agreement with BABAR's data is achieved with  $c=19.5 \pm 4.2$
$\left(18.9 \pm 4.1 \right)$~GeV$^{-4}$, $\chi_S=-19.4 \pm2.5
~(-19.8 \pm 2.6)$~GeV$^{-1}$, $N_P=1.122\pm0.034~(1.015 \pm 0.035)$
and $F^{BR_D}(m_\pi^2)=0.0977\pm0.0070$ (0.1010 $\pm$ 0.0072). 
The total $\chi^2$ is equal to  231.6 (233.5) for the 171 experimental points of the fit. 
For both fits the branching fraction for the $B^\pm \to \rho(770)^0 \pi^\pm,
\  \rho(770)^0 \to \pi^+ \pi^-$ decay is  $(8.1 \pm 0.5) \times 10^{-6}$, to be compared with 
the BABAR Collaboration determination of $(8.1\pm0.5\pm1.2^{+0.4}_{-1.1})\times 10^{-6}\approx (8.1 \pm 1.6)\times 10^{-6}$ from their isobar model analysis~\cite{Aubert:2009}.
Note that for the LO fit we explain essentially the BABAR Collaboration's result without significant 
modification of the $P$~wave normalization, the parameter $N_P\approx 1.02$ being close to 1.
For the NLO fit,  $N_P\approx1.12\pm0.03$ and one can compare $N_P^2-1\approx 25\%$ with the average
 20\% error of the experimental branching ratio.

The $CP$ average total branching fraction of the $B^\pm \to \pi^\pm \pi^\mp \pi^\pm$ decays calculated
 in the NLO fit is equal to $(15.2\pm 1.1)\times 10^{-6}$ to be compared to the measured value 
of $\left(15.2\pm0.6\pm1.2^{+0.4}_{-0.3}\right)\times 10^{-6}$ (table III of Ref.~\cite{Aubert:2009}).
The branching fraction for the $S$ wave equals to $(2.3 \pm 0.4)\times 10^{-6}$ and that for the
$D$ wave is $(2.8\pm 0.4)\times 10^{-6}$. 
The latter value is larger than the branching fraction for the 
$f_2(1270)\pi^{\pm}$, $(0.9 \pm 0.2 \pm 0.1^{+0.3}_{-0.1}) \times 10^{-6}$, determined in 
Ref.~\cite{Aubert:2009}.
In the experimental analysis the two resonances, namely $f_2(1270)$ and $f_0(1370)$, overlap to a large
extent, which makes their separation difficult and some part of the branching fraction obtained for one
resonance could have been attributed to the other one. 
The isobar model analysis of Ref.~\cite{Aubert:2009} gives $(2.9 \pm 0.5 \pm 0.5^{+0.7}_{-0.5}) \times 10^{-6}$
for the branching fraction of $f_0(1370)\pi^{\pm}$. Then, the sum of the branching
fractions for the two resonances equals to $3.8\times 10^{-6}$. 
This value compares well with the branching fraction
of $3.6\times 10^{-6}$ obtained by integrating our distribution in the $m_{\pi\pi}$ range 
between 1.0 and 1.64 GeV
in which both $f_2(1270)$ and $f_0(1370)$ give their dominant contributions.
In our model the $D$-wave contribution is dominant in this range.
Let us note that the value we obtain for the transition form factor $F^{BR_D}(m_\pi^2)$ is 29\%
larger than the
value 0.076 given in Table 1 of Ref.~\cite{Kim_PRD67_014002} for the ISGW2 model.
The $S$-wave contribution represents here as much as 15\% of the total branching fraction. 
This contribution is of the same order as that of the $\rho(1450)$ and $\rho(1700)$ 
which also represents 15~\% of the total $P$-wave contribution.

Before comparing our effective mass distributions to the experimental ones, we now give our result for the  pion scalar form factor  $\Gamma_1^n(s)$.
With  the fixed value of $\kappa=2$~GeV used in the fits, one obtains for the  $\alpha_i^n$, $\tau_i^n$ and $\omega_i^n$, $i=1,2,3$, entering into Eq.~(\ref{RjE}), the values given in Table~\ref{Tabalphai}.
Then, in Fig.~\ref{Gamma1N}, we show the modulus of the pion scalar form factor obtained using the NLO coefficients
 $a_i^p$ for the fitted value of the parameter $c= (19.5 \pm 4.2$ GeV)$^{-4}$ together with its envelope when
 $c$ varies within its error band. It is also compared to that of  the scalar form factor  calculated
 by Moussallam~\cite{PrivateBM} solving the Muskhelishvili-Omn\`es equations~[26] with a high-energy ansatz
 starting at 2~GeV and the same low-energy three coupled-channel scattering T-matrix as in our model
 (see Sec.~3.1).
However, in his calculation the off-diagonal matrix elements $T_{13}(E,k_i,p)$ and $T_{23}(k_i,E,p)$ are set to zero in the unphysical region $E<2m_3=1.4$~GeV.
Let us remind  here that the imaginary parts of these two pion form factors satisfy exactly the same relation given by Eq.~(\ref{Uni}).
The functional dependence of both $\Gamma_1^{n*}(s)$ moduli is quite similar.
It can be seen in Fig.~\ref{Gamma1N} that, within our model, the needed  $\Gamma_1^{n*}(s)$ is relatively
 well constrained.
 If we fix $\kappa=3$~GeV then the fit to BABAR data gives
 $c=(30.4 \pm 6.6) $~GeV${^{-4}}$, $\chi_S=(-20.2 \pm 2.9)$~GeV$^{-1}$ with a total $\chi^2$ of 234.1.
In this range of variation of the strongly correlated $\kappa$
 and $c$ parameters, we have checked that the scalar form factor varies smoothly. The corresponding
 values of the strength parameter $\chi_S$, being very close 
to $-20$~GeV$^{-4}$, are not sensitive to these variations.
 For $\kappa=3$~GeV the values of the branching 
fractions for 
the different $\pi\pi$ waves stay within the error bands of those for the $\kappa$ = 2~GeV case.

The threshold behavior of our pion form factor is governed by the chiral perturbation expansion Eq.~(\ref{GlowE}).
These ChPT constraints, not explicitly included in Moussallam's case, lead to $\Gamma_1^{n*}(s)$ moduli of both approaches to differ only slightly near the $\pi\pi$ threshold.
Above the $\pi \pi$ threshold, there is a maximum corresponding to the $f_0(600)$ resonance, then 
 close to 1~GeV a characteristic dip due to the $f_0(980)$ and finally, below the spike at 1.4 GeV 
related to the opening of the third channel, there is some enhancement generated by the $f_0(1400)$ present
 in the $\pi \pi$ three-channel model used here~\cite{Kaminski:1997gc, Kaminski_EPJC9_141}. 
The third threshold energy equal to 1.4 GeV is a parameter representing twice the mass of the effective two-pion mass $m_{(2\pi)}$ used to account for the four pion decays of scalar mesons (see Ref.~\cite{Kaminski:1997gc}). 
Thus, in nature there is no such sharp energy behavior.
These characteristic features of the pion scalar form factor $\Gamma_1^n(s)$ are essential to obtain a good fit of the experimental effective mass distributions of the $B^\pm$ to 3$\pi$ decays.

The results of the fit on the experimental distributions, obtained using the NLO coefficients $a_i^p$ 
in the $B^\pm \to \pi^\pm \pi^\mp \pi^\pm $ amplitudes, are displayed in 
Figs.~\ref{lightB-+}, \ref{lightB-cos-+} and \ref{lightB+cos-+}.
The $\rho(770)$-resonance contribution dominates  the $\pi^+\pi^-$ spectrum, but that of the $S$-wave 
is non negligible. 
As seen, the $S$-wave part is sizable near 500 MeV which is related to the contribution of the scalar 
resonance $f_0(600)$, not explicitely included in the BABAR Dalitz plot analysis~\cite{Aubert:2009}.
In the 1~GeV range the $f_0(980)$ resonance is not observed as a peak in the $\pi^+\pi^-$ spectrum. 
This fact is easily explained in our model since the decay amplitudes are proportional to the pion scalar form factor which has a dip near 1~GeV as seen in Fig.~\ref{Gamma1N}. 
Around 1.3 GeV there is a maximum coming from the contribution of the $f_2(1270)$ resonance.
Near 1.4 GeV the $f_0(1400)$ scalar resonance~\cite{Kaminski:1997gc, Kaminski_EPJC9_141} gives only
 a tiny enhancement in the distributions.  

Figure~\ref{lightB-+} exhibits a small $CP$ asymmetry, the $B^-$ and $B^+$ effective mass distributions
 being very close.
Summing the number of experimental events in the $m_{\pi^+\pi^-}$ range between 290 and 1640 MeV
 one finds 616 events for the $B^-$ decay and 606 for that of the $B^+$. This leads 
to a $CP$ asymmetry of ($0.8 \pm 4.8$)\% which can be compared to the values of 
$(1.7 \pm 0.2)\%$ for 
the NLO fit and $(-0.06\pm 0.08)\%$ for the LO fit.
Taking into account the statistical error of 4.8\% and adding to it a few percent systematic error one 
sees that both fits agree with experiment. 
Let us recall here the experimental value of the $CP$ asymmetry $A_{CP}=\left(3.2\pm4.4\pm3.1^{+2.5}_{-2.0}\right)$\% for the total sample of $\pi^{\pm}\pi^{\mp}\pi^{\pm}$ events~\cite{Aubert:2009}.
 For the particular decay mode, namely for the $B^{\pm}$ decay into
 $\rho(770)^0\pi^{\pm}$, $\rho(770)^0\rightarrow \pi^+\pi^-$, the isobar model analysis gives
$A_{CP}=\left(18\pm7\pm5^{+2}_{-14}\right)$\%, while from our model we get 
$3.6\% \pm 0.2 \%~(-0.03\% \pm 0.001\%)$.
Note here that the asymmetries obtained for the fit corresponding to the amplitudes calculated with the 
real LO $a_i^p$ coefficients are quite small as it could have been expected.

Figures~\ref{lightB-cos-+} and \ref{lightB+cos-+} show a spectacular feature, namely that the interference
 term of the $S$, $P$ and $D$ waves is quite important under the  $\rho(770)^0$ maximum. 
Here the $S$-$P$ interference dominates.  
The sign of this interference term depends on the sign of $\cos \theta$, so the $\rho$ peak is reduced for
 the negative values of $\cos \theta$ and enhanced for the positive values. 
This is a clear indication that the $\pi^+\pi^-$ effective mass distribution cannot be reproduced without 
the $S$-wave  contribution. If we try to fit the data without the $S$-wave amplitude then we obtain
a poor fit with $\chi^2 = 316.3$. In this case the effective mass distributions 
are not well described below 600~MeV and also under the $\rho$ maximum.
One striking feature is that the interference terms allow an extremely good representation of the separate $\cos \theta < 0$ and $\cos \theta > 0$ spectra for the $B^+$ decays (Fig.~\ref{lightB+cos-+}) and yield for the full spectrum [Fig.~\ref{lightB-+}b)] a $\chi^2$/point of 1.07.
The fit of the separate $B^-$ spectra (Fig.~\ref{lightB-cos-+}) is less satisfactory whereas that of the full spectrum [Fig.~\ref{lightB-+}a)] is almost perfect with a $\chi^2$/point of 1.2.

\section{Summary and outlook}
\label{Outlook}

The present paper is a continuation of our efforts~\cite{fkll,El-Bennich2006,Bppk, Leitner_PRD81} in constraining theoretically the meson-meson final state strong interactions in hadronic charmless three-body $B$ decays.
If the strong interaction amplitudes are sufficiently well understood then one can improve the precision of the weak interaction amplitudes extracted from these reactions.

Our theoretical model for the $B^\pm \to \pi^\pm \pi^\mp \pi^\pm$ is based on the application of the QCD 
factorization~\cite{Ali1998,Beneke:2001ev,bene03,Beneke2006} to quasi two-body processes in which only two
 of the three produced pions interact strongly, forming either an $S$-, $P$- or $D$-wave state.
One assumes that the third pion, being fast in  the $B$-meson decay frame, does not interact with this pair.
This hypothesis is mainly valid in a limited range of the $\pi^+ \pi^-$ effective mass, here taken between the $\pi \pi$ threshold and 1.64~GeV.

The short-distance interaction part of the decay amplitudes describes the flavor changing processes $b\to u \bar u d$  and $b \to d \bar d d$.
It is proportional to Cabibbo-Kobayashi-Maskawa matrix elements multiplied by effective coefficients calculable in the perturbative QCD formalism.
This short-distance amplitude is multiplied by a long-distance contribution expressed in terms of two products.
The first one is the product of the pion decay constant by the $B \to \pi \pi$ transition matrix element and the second one is the product of the pion form factor by the $B \to \pi$ transition form factor.
The parametrization [Eqs.~(\ref{F0pipi}),~(\ref{F1pipi})] of the scalar and vector $B$ to $\pi$ transition form factors follow from the light-cone sum rule study of Ref.~\cite{Ball_PRD71_014015}.

The effective Wilson coefficients are calculated to next-to-leading order in the strong coupling constant.
They include vertex and penguin corrections but neither hard-scattering ones nor annihilation contributions since these last two terms contain unknown phenomenological parameters related to amplitude divergences~\cite{bene03}.
We find that these vertex and penguin corrections are small in comparison to the leading order term (see Table~\ref{Tabailovp}).
However, they allow to generate some non-zero $CP$ asymmetries.

We then assume the $B$ to $\pi \pi$ transition matrix element to be equal to the product of the $B$ to 
intermediate meson transition form factor by the decay amplitude of this meson into two pions being either
in $S$, $P$ or $D$ wave.
The next step is to suppose the latter decay amplitude to be proportional to the pion non-strange scalar or vector form factor depending on the wave studied.
For the $S$ wave the proportionality factor is given by a fitted parameter $\chi_S$ and for the $P$ wave it is related to the inverse of the $\rho$ decay constant.
For the limited range of the effective $\pi \pi$ mass, from $\pi \pi$ threshold to $1.64$~GeV, 
the $B \to \pi \pi$ transition form factors are taken as constants given by the 
$B \to f_0(980)$~\cite{El-Bennich_PRD79} and by the $B \to \rho(770)$~\cite{Ball_PRD58_094016} 
transition form factors at $q^2=m_\pi^2$.
The decay amplitude for the $\pi\pi$ $D$ wave is described by a relativistic Breit-Wigner formula and the
not well known $B$ to $f_2(1270)$ transition form factor is fitted. We find $F^{Bf_2}(m_{\pi}^2)
=0.098 \pm 0.007$.

The pion scalar form factor is modeled by the unitary relativistic three coupled-channel 
equation~(\ref{gammai}) using  the $\pi \pi$, $K \bar K$ and effective $(2\pi)(2\pi)$ scattering $T$ matrix of Refs.~\cite{Kaminski:1997gc,Kaminski_EPJC9_141}.
This form factor depends on two fitted parameters: the first one $\kappa$ insures the convergence of the involved integrals and the second one, $c$, controls the high-energy behavior of the production functions accountable for the meson pair formation.
The pion vector form factor takes into account the contribution of the $\rho(770)$, $\rho(1450)$ and $\rho(1700)$, and follows from the parametrization of the Belle Collaboration in their study of the semi-leptonic $\tau^- \to \pi^- \pi^0 \nu_\tau$ decays.
For the $P$-wave amplitude we introduce a fitted correction factor $N_P$. 

We obtain a good fit to the $\pi \pi$ effective mass distributions of the BABAR Collaboration data of the
 $B^\pm \to \pi^\pm \pi^\mp \pi^\pm$  decays~\cite{Aubert:2009}.
The value of the branching fraction for the 
$B^\pm \to \rho(770)^0 \pi^\pm$ decays, $(8.1\pm0.7\pm1.2^{+0.4}_{-1.1})\times 10^{-6}$,
is well reproduced with the correction factor $N_P$ close to 1. This shows that the QCD 
factorization gives the right strength of the $B$ to $\rho\pi$ decay amplitude. 
The $\pi^+ \pi^-$ spectra are dominated by the $\rho(770)^0$ resonance but, at low effective mass, 
the $S$-wave contribution is sizable.
Here the $f_0(600)$ resonance manifests its presence.
Furthermore one observes a strong interference of the $S$ and $P$ waves in the event distributions for 
$\cos \theta >0$ and $\cos \theta <0$.
Here the $f_0(980)$ is not directly visible as a peak, since the pion scalar form factor has a dip near
1~GeV.
The surplus of events in the $\pi^+\pi^-$ effective mass close to 1.25 GeV is well described by the 
contribution of the $f_2(1270)$ resonance. The branching fraction for $B^\pm \to f_2(1270) \pi^\pm,
\  f_2(1270) \to \pi^+ \pi^-$ decay is found to be of $(2.8 \pm 0.4) \times 10^{-6}$.
At 1.4~GeV, the tiny maximum of the $S$-wave distribution comes from the scalar resonance
 $f_0(1400)$~\cite{Kaminski:1997gc,Kaminski_EPJC9_141}.

Our model yields a unified description of the contribution of the three scalar resonances
 $f_0(600)$, $f_0(980)$ and $f_0(1400)$ in terms of one function: the pion non-strange scalar form factor.
This reduces strongly the number of needed free parameters to analyze the Dalitz plot.
The functional form of our $S$-wave amplitude [Eq.~(\ref{MSsij})], proportional to $\Gamma _1^{n*}(s)$,
 could be used in Dalitz-plot analyses and the table of  $\Gamma _1^{n*}(s)$ values can be sent upon 
request.

The strong interaction phases of the decay amplitudes are constrained by unitarity and meson-meson data.
Their determination should help in the extraction of the weak angle phase $\gamma$ or $\phi_3$ equal to
 $\arg(-\lambda_u^*/\lambda_c^*)$.
Of course new experimental data with better statistics would be welcome.
One expects $B^\pm \to \pi^\pm \pi^\mp \pi^\pm$  events from the Belle Collaboration, and probably, in the
near future, from LHCb and from the near term super $B$ factories.

\vspace*{6pt}
The authors are obliged to Bachir Moussallam for providing them the values of his pion scalar form factor $\Gamma_1^n(s)$ and to Gagan Bihari Mohanty for useful comments on the BABAR data.
We are very grateful to Maria R\'o\.{z}a\'nska, Bachir Moussallam, Eli Ben-Haim and Jos\'e Ocariz for helpful discussions.
This work has been supported in part by the Polish Ministry of Science and
Higher Education (grant No N N202 248135) and by the IN2P3-Polish Laboratories
Convention (project No 08-127). 

\appendix
\section{\bf{Long-distance functions $X_{S,P,D}$ and $Y_{S,P}$}}
\label{appendix}

\subsection{\bf{~The function $X_S$ from the $S$-wave amplitude proportional to $BR_S$ transition
 matrix element}}
\label{SwaveF0BRS}

From Eq.~(\ref{XS}) the function $X_S$ reads 
\begin{eqnarray}
\label{X_S}
X_{{S}} &\equiv& \left \langle  [\pi^+(p_2){\pi^-}(p_3) ]_{S} \vert (\bar u b)_{V-A}\vert B^-\right \rangle \ho \left \langle \pi^-\vert  (\bar d u)_{V-A}\vert 0 \right \rangle \nonumber\\
&=& G_{R_S \pi^+\pi^-}^n(s_{23}) \left \langle  R_{S} \vert (\bar u b)_{V-A}\vert B^-\right \rangle \ho \left \langle \pi^-\vert  (\bar d u)_{V-A}\vert 0 \right \rangle, 
\end{eqnarray}
where the vertex function  $G_{R_S \pi^+\pi^-}^n(s_{23})$ describes the $R_S$ decay into a $[\pi^+\pi^-]_S$ pair.
The $B$ to $R_S$ transition matrix element reads (see e.g. Eq.~(B6) of Ref.~\cite{Cheng0704.1049})

\begin{multline}
\label{ffRSB}
\langle R_S(p_{2}+p_{3})\vert \bar u\gamma^\mu(1-\gamma^5)b\vert B^-(p_{B})\rangle=\\
i\Bigg\{\left[
(p_{B}+p_{2}+p_{3})^\mu-\dfrac{M_{B}^2-s_{23}}{m_\pi^2}p_1^\mu
\right]F_1^{B R_S}(m_\pi^2)
+ \dfrac{M_{B}^2-s_{23}}{m_\pi^2}p_1^\mu F_0^{B R_S}(m_\pi^2)\Bigg\},
\end{multline}
where $F_{0}^{B R_S}(m_\pi^2)$ and   $F_{1}^{B R_S}(m_\pi^2)$ are the $B R_S$  scalar and vector form factors, respectively.
The pion decay constant $f_\pi$ is defined as

\begin{equation}
\label{fpi}
\langle \pi^-(p_1)\vert \bar d\gamma_\mu(1-\gamma_5) u\vert 0 \rangle = i f_\pi p_{1\mu}.
\end{equation}
The product of Eqs.~(\ref{ffRSB}) and (\ref{fpi}) yields 

\begin{equation}
\label{RSBfpi}
  \left \langle R_{S}\vert (\bar u b)_{V-A}\vert B^-\right \rangle 
 \left \langle \pi^-\vert  (\bar d u)_{V-A}\vert 0 \right \rangle =  -(M_{B}^2-s_{23}) f_\pi F_0^{B R_S}(m_\pi^2).
\end{equation}
The vertex function $G_{R_S \pi^+\pi^-}^n(s_{23})$, as in Ref.~\cite{El-Bennich2006}, is modeled by

\begin{equation}
\label{vertex}
\left \langle [\pi^+\pi^-]_{S} \vert \bar nn\vert 0\right \rangle=G_{R_S\pi^+\pi^-}^n(s_{23})\left \langle R_S\vert \bar nn\vert 0\right\rangle.
\end{equation}
An effective scalar decay constant $f_{R_S}^n$ can be introduced with

\begin{equation}
\label{RSdecay}
\langle R_S \vert \bar nn\vert 0\rangle=m_{R_S} \ho f_{R_S}^n.
\end{equation}
From Eqs.~(\ref{vertex}), (\ref{gamman}) and (\ref{RSdecay}) one obtains

\begin{equation}
\label{RSvertex}
G_{R_S\pi^+\pi^-}^n(s_{23})=\sqrt{\frac{2}{3}}\ho \chi_S \ho \Gamma_1^{n*}(s_{23})=\sqrt{\frac{2}{3}}\ho \frac{\sqrt{2}B_0}{m_{R_S}\ho f_{R_S}^n}\Gamma_1^{n*}(s_{23}),
\end{equation}
with

\begin{equation}
\label{chi}
\chi_S= \frac{\sqrt{2}B_0}{m_{R_S}\ho f_{R_S}^n}.
\end{equation}
The effective scalar decay constant has a role comparable to the $R_P$ decay constant as can be seen comparing Eqs.~(\ref{RSvertex}) and (\ref{vertexRP}).
The product of Eqs.~(\ref{RSvertex}), (\ref{ffRSB}) and (\ref{fpi}) gives
\begin{equation}
\label{chiRS}
X_{S} = - \sqrt{\frac{2}{3}} \ho \chi_S \ho f_\pi \ho (M_{B}^2-s_{23}) \ho F_0^{BR_S}(m_\pi^2) \ho\Gamma_1^{n*}(s_{23}).
\end{equation}

\subsection{\bf{~The function $Y_S$ from the $S$-wave amplitude proportional to $B\pi$ transition matrix element}
\label{SwaveF0Bpi}}

From Eq.~(\ref{ZS}) one has
\begin{eqnarray}
 \label{Y_S}
Y_{{S}} &\equiv& \left \langle \pi^-\vert (\bar d b)_{sc-ps}\vert B^- \right \rangle \ho \left \langle  [\pi^+(p_2){\pi^-}(p_3) ]_{S}\vert (\bar d d)_{sc+ps}\vert 0\right \rangle\nonumber\\
&=& \left \langle \pi^-\vert \bar d b \vert B^- \right \rangle \left\langle  [\pi^+(p_2){\pi^-}(p_3) ]_{S}\vert \bar d d\vert 0\right \rangle.
\end{eqnarray}
From the Dirac equations satisfied by $b(p_B)$ and $\bar d (p_1)$  one obtains

\begin{equation}
\label{Bpidbarb}
 \left \langle \pi^-(p_1) \left \vert  \bar d(p_1) b(p_B) \right \vert B^-(p_B) \right \rangle=
 \left \langle \pi^-(p_1 )\left \vert  \bar d(p_1) \frac{\gamma \cdot\left(p_B-p_1\right)}{m_b-m_d} b(p_B) \right \vert B^-(p_B) \right \rangle.
\end{equation}
The $B$ to $\pi$ transition matrix element $\left \langle \pi^-\vert  (\bar d b)_{V-A}\vert B^- \right \rangle$, entering into the above expression, can be written as (see e.g. Eq.~(5) of Ref.~\cite{Bppk})

\begin{multline}
\label{ffpiB}
\langle \pi^-(p_{1})\vert \bar d\gamma^\mu(1-\gamma^5)b\vert B^-(p_{B})\rangle\\
=\left[
(p_{B}+p_{1})^\mu-\dfrac{M_{B}^2-m_{\pi}^2}{q^2}q^\mu
\right] F_1^{B\pi}(q^2)
+\dfrac{M_{B}^2-m_{\pi}^2}{q^2}q^\mu F_0^{B\pi}(q^2) ,
\end{multline}
where $F_{0}^{B\pi}(q^2)$ and $F_{1}^{B\pi}(q^2)$ are the $B\pi $ scalar and vector 
form factors, respectively and $q=p_{B}-p_{1}=p_{2}+p_{3}$.
Using Eqs.~(\ref{ffpiB}) and (\ref{gamman}) in Eq.~(\ref{Y_S}), yields 
\begin{equation}
\label{Heff3}
Y_{S} =  \sqrt{\frac{2}{3}} \ho B_0\ho \Gamma_1^{n*}(s_{23})\ho \frac {M_B^2-m_\pi^2}{m_b-m_d}
\ho F_0^{B\pi}(s_{23}).
\end{equation}

\subsection{\bf{~The function $X_P$ from the $P$-wave amplitude proportional to $BR_P$ transition matrix element}
\label{PwaveF1BRp}}

From Eq.~(\ref{XP}) one has for the function $X_P$ (see Eq.~(3.1) of Ref.~\cite {Cheng0704.1049})
\begin{eqnarray}
\label{X_P}
X_{{P}} 
\equiv
 \left \langle  [\pi^+(p_2){\pi^-}(p_3) ]_{P} \vert (\bar u b)_{V-A}\vert B^-\right \rangle \ho \left \langle \pi^-\vert  (\bar d u)_{V-A}\vert 0 \right \rangle \nonumber\\ 
=
 \frac{G_{R_P \pi^+\pi^-}^n(s_{23})} {\sqrt{2}}\ \epsilon \cdot(p_2-p_3) 
\left \langle  R_{P} \vert (\bar u b)_{V-A}\vert B^-\right \rangle \ho
\left \langle \pi^-\vert  (\bar d u)_{V-A}\vert 0 \right \rangle, 
\end{eqnarray}
where the $R_P$ decay into a $[\pi^+\pi^-]_P$ pair is described by
the vertex function  $G_{R_P\pi^+\pi^-}^n(s_{23})$.
Here $\epsilon$ represents the polarization vector of the $P$-wave meson $R_P$.
 The factor $1/\sqrt{2}$ comes from the fact that $R_P$ represents the $\rho(770)^0$.   As seen from e.g. Eq.~(B6) of Ref.~\cite{Cheng0704.1049} or Eq.~(24) of Ref.~\cite{Ali1998},
 
\begin{eqnarray}
\label{RPBFF}
  \left \langle R_{P}(p_2+p_3)\vert (\bar u b)_{V-A}\vert B^-(p_B)\right \rangle=
   -i \ho 2m_{R_P} \ho \frac {\epsilon^* \cdot p_B}{p_1^2}\ho p_1\ho A_0^{B R_P}(p_1^2)
\nonumber\\ 
+ {\rm other\ terms}.
\end{eqnarray}
The ``other terms" do not give any contribution when multiplying this matrix element by that given in Eq.~(\ref{fpi}).
Plugging this expression into  Eq.~(\ref{X_P}) one has
a product of polarization vectors and the sum over the three possible polarization eigenvalues
of the state $R_P$ should be done. From

\begin{equation}
\label{sumepsilon}
\sum_{\lambda=0,\pm 1} \epsilon_\mu^\lambda(p) \epsilon_\nu^{\lambda *}(p)=
-(g_{\mu\nu}-\frac{p_\mu p_\nu} {p^2}),
\end{equation}
one obtains  
\begin{equation}
\label{ }
 \sum_{\lambda=0,\pm 1} \epsilon^{\lambda}\cdot(p_2-p_3) \epsilon^{\lambda^*} \cdot p_B=-p_1\cdot (p_2-p_3).
\end{equation}
Then
\begin{equation}
\label{pi+pi-P1}
X_{P}=  N_P \frac{f_\pi}{f_{R_P}}\ho (s_{13} - s_{12}) \ho
A_0^{B R_P}(m_\pi^2)\ho F_1^{\pi\pi}(s_{23}). 
\end{equation}
Above, as shown in Ref.~\cite{Bppk} for the $K^*(892) \to (K \pi)_P$ decay case [see their Eq.~(D9)], we have parametrized the $R_P \pi^+\pi^-$ vertex function in terms of the pion vector form factor $ F_1^{\pi\pi}(s_{23})$.
One has
\begin{equation}
\label{vertexRP}
 G_{R_P\pi^+\pi^-}(s_{23})=N_P \frac{\sqrt{2}}{m_{R_P} f_{R_P}} F_1^{\pi\pi}(s_{23}),
\end{equation}
$f_{R_P}$ being the charged $R_P$ decay constant.
Above we have introduced a parameter $N_P$ to take into account the possible deviation of the strength of the $P$ wave, here proportional to $1/f_{R_P}$. 

\subsection{\bf{~The function $Y_P$ from the $P$-wave amplitude proportional to the $B\pi$ transition 
matrix element}}
\label{PwaveF1Bpi}
From Eq.~(\ref{YP})
\begin{equation}
 \label{Y_P}
Y_{{P}} \equiv \left \langle \pi^-\vert  (\bar d b)_{V-A}\vert B^- \right \rangle \ho \left \langle  [\pi^+(p_2){\pi^-}(p_3) ]_{P}\vert (\bar u u)_{V-A} \vert 0 \right \rangle. 
\end{equation}

The pion vector form factor is defined by (see e.g. Eq.~(36) of Ref.~\cite{Ali1998})

\begin{equation}
\label{ffpipiP}
\langle R_P\vert (\bar u u)_{V-A}\vert 0 \rangle = \langle [\pi^+(p_{2}) \pi^-(p_{3})]_P\vert \bar u\gamma_\mu(1-\gamma_5)u\vert 0\rangle
=-\left (p_{2}-p_{3}\right)_\mu F_1^{\pi \pi}(q^2).
\end{equation}
The minus sign arises from the definition of the form factor $F_1^{\pi \pi}(q^2)$ which contains a plus sign for a $(\bar d d)_{V-A}$ current [similar to Eq.~(\ref{ffpiB}], then as $\rho^0= 1/\sqrt{2} (u\bar u -d \bar d)$, there will be a minus sign for a  $(\bar u u)_{V-A}$ current.
The product of Eqs.~(\ref{ffpiB}) and (\ref{ffpipiP}) gives
\begin{equation}
\label{ffpiBxpipiP }
Y_{P} = -2 \ho p_1 \cdot \left (p_{2}-p_{3}\right)  F_1^{B \pi}(q^2) F_1^{\pi \pi}(q^2)=  (s_{13} - s_{12}) F_1^{B \pi}(q^2) F_1^{\pi \pi}(q^2).
\end{equation}
 
\subsection{\bf{~The function $X_D$ from the $D$-wave amplitude proportional to $BR_D$ transition
 matrix element}}
\label{DwaveFBRD}

From Eq.~(\ref{XD}) one has

\begin{eqnarray}
\label{X_D}
X_{{D}} 
\equiv 
 \left \langle  [\pi^+(p_2){\pi^-}(p_3) ]_{D} 
 \vert (\bar u b)_{V-A}\vert B^-\right \rangle \ho \left 
\langle \pi^-(p_1)\vert  (\bar d u)_{V-A}\vert 0 \right \rangle \nonumber \\ 
=
\frac{1}{\sqrt2} G_{R_D \pi^+\pi^-}(s_{23}) \sum_{\lambda=-2}^2 \epsilon_{\alpha\beta}
p_2^{\alpha} p_3^{\beta}
\left \langle  R_{D}^\lambda(p_D) \vert (\bar u b)_{V-A}\vert B^-\right \rangle \ho
 \nonumber\\
\left \langle \pi^-(p_1)\vert  (\bar d u)_{V-A}\vert 0 \right \rangle, 
\end{eqnarray}
with $p_D=p_2+p_3$. 
 The factor of $1/\sqrt2$ is due to the quark content of the resonance $R_D$ [the meson 
$f_2(1270)$]. 
The $R_D$ decay into a $[\pi^+\pi^-]_D$ pair is described by
 the vertex function $G_{R_D \pi^+\pi^-}(s_{23})$. 
 Here  $\epsilon_{\alpha\beta}(\lambda)$ represents the polarization tensor of the $f_2(1270)$ and 
 $\lambda $ is its spin projection (see  Ref.~\cite{Pilkuhn1967}, p.~147). 
Taking Eq.~(A3) for  $\left \langle \pi^-(p_1)\vert  (\bar d u)_{V-A}\vert 0 \right \rangle$
and Eq.~(4) of Ref.~\cite{Kim_PRD67_014002} for the transition matrix element 
$\left \langle  R_{D}^{\lambda}(p_D)\vert (\bar u b)_{V-A}\vert B^-\right \rangle $ we obtain

\begin{equation}
\label{XDE}
X_{{D}}=-\frac{f_{\pi}}{\sqrt{2}}G_{R_D \pi^+\pi^-}(s_{23}) F^{BR_D}(m_{\pi}^2)
\sum_{\lambda=-2}^2 \epsilon_{\alpha\beta}(\lambda)p_2^{\alpha} p_3^{\beta}
\epsilon_{\mu\nu}^*(\lambda)p_B^{\nu} p_1^{\mu}.
\end{equation}
To be consistent with the choice of normalization of Eq. (A2), we have multiplied by $i$ the right hand side of Eq. (4) in Ref.~\cite{Kim_PRD67_014002}.
One can show that (see Eqs. (7.7) and (7.8) of  Ref.~\cite{Pilkuhn1967}, p.~73)

\begin{equation}
\label{De}
D(s_{12},s_{23}) \equiv \sum_{\lambda=-2}^2 \epsilon_{\alpha\beta}(\lambda)p_2^{\alpha} p_3^{\beta}
\epsilon_{\mu\nu}^*(\lambda)p_B^{\nu} p_1^{\mu}
=\frac{1}{3}(\vert \overrightarrow{p_1} \vert \vert \overrightarrow{p_2} \vert )^2
-(\overrightarrow{p_1}\cdot \overrightarrow{p_2})^2,
 \end{equation}
$\overrightarrow{p_1}$ and $\overrightarrow{p_2}$  being the momenta of the $\pi^-(p_1)$ and 
the $\pi^+(p_2)$ in the rest frame of $\pi^+(p_2)$ and $\pi^-(p_3)$. One obtains
, with $m_{23}=\sqrt{s_{23}}$,

\begin{eqnarray}
\label{kincms23}
\overrightarrow{p_1}\cdot \overrightarrow{p_2}&=&\frac{1}{4}(s_{13}-s_{12}), \nonumber \\
\vert \overrightarrow{p_2} \vert& = & \frac{1}{2}\sqrt{m_{23}^2-4m_\pi^2},
\hspace{1cm} \vert \overrightarrow{p_2} \vert = \vert \overrightarrow{p_3} \vert,  \nonumber \\
\vert \overrightarrow{p_1} \vert  & = &  \frac{1}{2m_{23}}\sqrt{\left[M_B^2-\left(m_{23}+m_\pi\right)^2\right]
\left[M_B^2-\left(m_{23}-m_\pi\right)^2\right]}, 
\end{eqnarray}
which allows to express Eq.~(\ref{De}) in terms of $s_{12}$ and $s_{23}$. 
The vertex function entering into Eq.~(\ref{X_D}) is parametrized as being proportional to a relativistic
  Breit-Wigner resonance formula, we write

\begin{equation}
\label{GRD}
G_{R_D \pi^+\pi^-}(s_{23})= \sqrt{\frac{2}{3}}\frac{G_{f_2}}{m^2_{R_D}-s_{23}-im_{R_D}\Gamma(s_{23})},
\end{equation}
where  (see  Ref.~\cite{Pilkuhn1967}, p.147)
\begin{equation}
\label{Gf2}
G_{f_2}= m_{f_2}\sqrt{\frac{60\pi\Gamma_{f_2\pi\pi}}{q_{f_2}^5}},\hspace{1cm}
\Gamma_{f_2\pi\pi}=0.848~\Gamma_{f_2}
\end{equation}
and the mass-dependent width $\Gamma(s_{23})$ can be expressed as
 (see Eq.~(7) of Ref.~\cite{Aubert:2009}),

\begin{equation}
\label{Gamf2}
\Gamma(s_{23}) = \Gamma_{f_2} \left (\frac{\vert \overrightarrow{p_2} \vert }{q_{f_2}} \right 
)^5 \frac{m_{f_2}}{m_{23}}
 \frac{X(\vert \overrightarrow{p_2} \vert )}{X(q_{f_2})}.
\end{equation}
Here $\Gamma_{f_2} $ is the total width of the $f_2(1270)$ resonance, $m_{f_2}$ its mass and
$q_{f_2}$ is the pion momentum in the $f_2$ c.m. system. The Blatt-Weisskopf barrier form factor is given by~\cite{Aubert:2009}

\begin{equation}
\label{Xz}
X(z) = \frac{1}{(zr_{BW})^4+3(zr_{BW})^2+9},
\end{equation}
where the meson radius parameter $r_{BW}=4$ (GeV/c)$^{-1}$.
Finally one has
\begin{equation}
\label{XDB}
X_D= -\frac{1}{\sqrt2}f_\pi   F^{B R_D}(m_\pi^2)  \ho
\sqrt{\frac{2}{3}}\frac{G_{f_2} D(s_{12},s_{23})}{m^2_{R_D}-s_{23}-im_{R_D}\Gamma(s_{23})}.
\end{equation}

\section{\bf {Linear system of equations for $\alpha_i^n$, $\tau_i^n$ and $\omega_i^n$}}
\label{alpha_i_equation}
The linear system of nine equations satisfied by the nine  production function parameters $\alpha_i^n$, $\tau_i^n$ and $\omega_i^n$, $i=1,2,3$, is

\begin{eqnarray}
\label{alphatauomega}
 \alpha_i^{n} + \sum_{j=1}^3 \alpha_j^{n} H_{ji}(0)& = & d_i^{n}, \nonumber\\
\tau_i^{n} + \sum_{j=1}^3 \left(\tau_j^{n} H_{ji}(0)+\alpha_j^{n} \frac {\partial H_{ji}(E)}{\partial E} \bigg\vert_{E=0}\right )& = & 0, \nonumber\\
\omega_i^{n} + \sum_{j=1}^3 \left(\omega_j^{n} H_{ji}(0)+\tau_j^{n} \frac {\partial H_{ji}(E)}{\partial E} \bigg\vert_{E=0}+\frac{1}{2}\alpha_j^{n}
 \frac {\partial^2 H_{ji}(E)}{\partial E^2} \bigg\vert_{E=0}\right )
=f_i^{n}.
\end{eqnarray}


\begin{thebibliography}{99}

 \bibitem{fkll} 
A. Furman, R. Kami\'nski, L.~Le\'sniak and B.~Loiseau,  
Phys. Lett.  B {\bf 622}, 207 (2005),
\textit{Long-distance effects and final state interactions in $B \to \pi\pi K$ and $B \to K\bar K K$ decays}.

 \bibitem{El-Bennich2006}
   B. El-Bennich, A. Furman, R. Kami\'nski, L.~Le\'sniak and B.~Loiseau,
   Phys. Rev. D \textbf{74}, 114009 (2006),
   \textit{Interference between $f_0(980)$ and $\rho(770)^0$ resonances in $B\to\pi^+\pi^-K$ decays}.

\bibitem{Bppk}
 B. El-Bennich, A.~Furman, R.~Kami\'nski, L.~Le\'sniak,  B.~Loiseau, B.~Moussallam,  Phys. Rev. D {\bf 79}, 094005 (2009), \textit{$CP$ violation and kaon-pion interactions in $B \to K \pi^+ \pi^- $ decays}.

\bibitem{Leitner_PRD81}
O. Leitner, J.-P. Dedonder, B. Loiseau, and R. Kami\'nski,
Phys. Rev. D {\bf 81}, 094033 (2010),
 \textit{$K^*$ resonance effects on direct $CP$ violation in $B \to\pi \pi K$}.
 
\bibitem{Aubert:2009}
B. Aubert, \textsl{et al.} (BABAR Collaboration),
Phys. Rev. D \textbf{79}, 072006 (2009),
 \textit{Dalitz-plot analysis of $B^\pm \to \pi^\pm \pi^\mp \pi^\pm$ decays.}

\bibitem{Ali1998}
A. Ali, G. Kramer  and Cai-Dian L\"u,
Phys. Rev. D \textbf{58}, 094009 (1998),
\textit{Experimental tests of factorization in charmless nonleptonic two-body $B$ decays}.


\bibitem{Beneke:2001ev}
  M.~Beneke, G.~Buchalla, M.~Neubert and C.~T.~Sachrajda,  Nucl.\ Phys.\  {\bf B606}, 245 (2001), \textit{QCD factorization in $B \to \pi K, \pi \pi$ decays and extraction of Wolfenstein  parameters}.

\bibitem {Gardner_PRD65}
S.~Gardner and U.-G.~Mei\ss ner, 
Phys. Rev. D \textbf{65},  094004 (2002),
\textit{Rescattering and chiral dynamics in $B \to\rho \pi$ decays}.

\bibitem{bene03}
  M.~Beneke and M.~Neubert,
  Nucl. Phys. \textbf{B675}, 333 (2003), 
  \textit{QCD factorization for $B\to PP$ and $B\to PV$ decays}.

\bibitem {LeitnerQCDF}
 O. Leitner, X-H. Guo, A.W. Thomas,
J. Phys. G: Nucl. Part. Phys. \textbf{31}, 199 (2005),
\textit{Direct $CP$ violation, branching ratios and form factors $B\to\pi,\ B\to K$ in $B$ decays.}

 \bibitem{Cheng:2005nb}
  H.~Y.~Cheng, C.~K.~Chua and K.~C.~Yang,
 Phys.\ Rev.\  D {\bf 73}, 014017  (2006),
  \textit{Charmless hadronic B decays involving scalar mesons: implications to  the  nature of light scalar mesons.}

\bibitem {Cheng0704.1049} 
H-Y. Cheng, C-K. Chua and A. Soni, Phys. Rev D \textbf{76}, 094006 (2007), \textit{Charmless three-body decays of B mesons}.

\bibitem{Beneke2006}
M. Beneke, Nucl. Phys. B (Proc. Suppl.) \textbf{170}, 57 (2007),
\textit{Hadronic B decays}.

\bibitem {Deandrea_PRL86}
A.~Deandrea and A. D.~Polosa, 
Phys. Rev. Lett. \textbf{86},  216 (2001),
\textit{$B \to\rho \pi$ decays, Resonant and Nonresonant Contributions}.

\bibitem{BABAR_PRD76}
B. Aubert, \textsl{et al.} (BABAR Collaboration),
Phys. Rev. D \textbf{76}, 012004 (2007),
 \textit{Measurement of $CP$-violating asymmetries in $B^0 \to (\rho \pi)^0$ using a time-dependent  Dalitz-plot analysis.}

\bibitem{Belle_PRD77}
A. Kusaka, \textsl{et al.} (Belle Collaboration),
Phys. Rev. D \textbf{77}, 072001 (2008),
 \textit{Measurement of $CP$ asymmetries and branching fractions in a time-dependent  Dalitz-plot analysis of $B^0 \to (\rho \pi)^0$ and a constraint on the quark mixing angle $\phi_2$.}

\bibitem{Beneke3body}
M. Beneke in Three-Body Charmless B Decays Workshop,\\
 \texttt{http://lpnhe-babar.in2p3.fr/3BodyCharmlessWS/}, February 1-3, 2006, LPNHE, Paris,
\textit{Quasi two-body and three-body decays in the heavy quark expansion}.
    

\bibitem{PDG}
 K. Nakamura {\it et al.} (Particle Data Group),
  J. Phys. G {\bf 37} 075021 (2010), \textit{Review of particle physics}.

\bibitem{El-Bennich_PRD79}
B. El-Bennich, O. Leitner, J.-P. Dedonder, B. Loiseau,
Phys. Rev. D {\bf 79}, 076004 (2009), 
\textit{The Scalar Meson $f_0(980)$ in Heavy-Meson Decays}.

\bibitem{Ball_PRD58_094016}
P. Ball, V. M. Braun,
Phys. Rev. D {\bf 58}, 094016 (1998),
\textit{Exclusive semileptonic and rare $B$ meson decays in QCD}.

\bibitem{Ball_PRD71_014015}
P. Ball and R. Zwicky
Phys. Rev. D {\bf 71}, 014015 (2005),
\textit{New results on $B \to \pi,\ K,\eta$ decay form factors from light-cone sum rules}.

\bibitem {meis01}
U.-G.~Mei\ss ner and J.~A.~Oller, Nucl. Phys. \textbf{A679}, 671 (2001),
 \textit{$J/Psi \to \phi \pi \pi (K\bar K)$ decays, chiral dynamics and OZI violation.}

\bibitem{Kim_PRD67_014002}
C. S. Kim, Jong-Phil Lee, and Sechul Oh,
Phys. Rev. D {\bf 67}, 014002 (2003) [8 pages]
\textit{Nonleptonic two-body charmless $B$ decays involving a tensor meson in the ISGW2 model}.

\bibitem{Barton65} 
G. Barton, \textit{Introduction to dispersion techniques in field theory}, Benjamin, New-York, 1965.

\bibitem{Cheng_PRD69} 
H.-Y. Cheng, C. K. Chua and C. W. Hwang,
Phys Rev. {\bf D} 69, 074025 (2004),
\textit{Covariant light-front approach for $S$-wave and $P$-wave mesons: Its application to decay constants and form factors}.

\bibitem{MO}
N. I. Muskhelishvili, Singular integral equations, (P.Nordhof 1953), chapters 18 and 19;
R. Omn\`es,
Nuovo Cim. {\bf 8}, 316 (1958),
\textit{On the Solution of certain singular integral equations of quantum field theory.}

\bibitem{DonoghueNPB343} 
 J. F. Donoghue, J. Gasser, H. Leutwyler,
 Nucl. Phys. {\bf 343}, 341 (1990),
\textit{The decay of a light Higgs boson}.

\bibitem{L6}
  B.~Moussallam,
  Eur.\ Phys.\ J.\  C {\bf 14}, 111 (2000),
  \textit{$N_f$ dependence of the quark condensate from a chiral sum rule}.
 
\bibitem{Kaminski:1997gc}
  R.~Kami\'nski, L.~Le\'sniak and B.~Loiseau,
  Phys. Lett. B {\bf 413} (1997) 130,
 \textit{Three channel model of meson meson scattering and scalar meson spectroscopy}.

 \bibitem{Kaminski_EPJC9_141}
  R.~Kami\'nski, L.~Le\'sniak and B.~Loiseau,
 Eur. Phys. J. {\bf C9}, 141 (1999),
 \textit{Scalar mesons and multichannel amplitudes.}

\bibitem{KLM_PRD50_3145}
R.~Kami\'nski and L.~Le\'sniak, J.-P. Maillet,
Phys. Rev. D {\bf 50}, 3145 (1994), 
\textit{Relativistic effects in scalar meson dynamics.}

\bibitem{Yamaguchi}
 Yoshio Yamaguchi and Yoriko Yamaguchi ,
 Phys. Rev. {\bf 95}, 1635 (1954), 
 \textit{Two-Nucleon Problem When the Potential Is Nonlocal but Separable. II}.

\bibitem{LM_PRD74_034021}
T. A. L\"ahde, and U.-G. Mei\ss ner,
Phys. Rev. D {\bf 74}, 034021 (2006), 
\textit{Improved analysis of $J/\psi$ decays into a vector meson and two pseudoscalars}.

\bibitem{Alton_PRD78_114509}
C. Allton,  \textsl{et al.}  (RBC and UKQCD Collaborations), 
Phys. Rev. D {\bf 78}, 114509 (2008),
\textit{Physical results from $2+1$ flavor domain wall QCD and SU(2) chiral perturbation theory}.

\bibitem{Fujikawa_PRD78_072006}
M. Fujikawa {\it et al.} (Belle Collaboration),
Phys. Rev. D {\bf 78}, 072006 (2008),
\textit{High-statistics study of the $\tau^-\to\pi^-\pi^0\nu_\tau$ decay}.

\bibitem{BallJones2007}
P. Ball and G. W. Jones,
JHEP 0703, 69 (2007),
\textit{Twist-3 distribution amplitudes of K* and $\phi$ mesons}.

\bibitem{PrivateBM}
B. Moussallam, private communication.

\bibitem{Pilkuhn1967}
H. Pilkuhn, \textit{The interactions of hadrons}, North-Holland P. C., 1967.

\end{thebibliography}
\end{document}